%% file: ZrS2.tex
\documentclass[prl,twocolumn,notitlepage,superscriptaddress]{revtex4-1}
\usepackage{amsmath}
\usepackage{amssymb}
\usepackage{wasysym}
\usepackage{graphicx}
\usepackage{color}
\usepackage[caption=false]{subfig}

\renewcommand{\k}{\mathbf{k}}
\newcommand{\q}{\mathbf{q}}

\newcommand{\ND}[1]{\hat{n}_{#1}}

\renewcommand{\r}{\mathbf{r}}

\newcommand{\tauID}{\hat{\mathbf{1}}}
\newcommand{\tauX}{\hat{\boldsymbol{\tau}}_x}
\newcommand{\tauY}{\hat{\boldsymbol{\tau}}_y}
\newcommand{\tauZ}{\hat{\boldsymbol{\tau}}_z}

\newcommand{\captiontitle}[1]{\textbf{#1}}

\input{macros.tex}

\begin{document}
\title{Ultra-Strong Spin-Orbit Coupling and Topological Moir\'e Engineering in Twisted ZrS$_2$ Bilayers}

\author{Martin Claassen}
\altaffiliation{M.C. and L.X. contributed equally to this paper.}
\affiliation{Department of Physics and Astronomy, University of Pennsylvania, Philadelphia, PA 19104}
\altaffiliation{claassen@sas.upenn.edu}

\author{Lede Xian}
\altaffiliation{M.C. and L.X. contributed equally to this paper.}
\affiliation{Songshan Lake Materials Laboratory, 523808 Dongguan, Guangdong, China}
\altaffiliation{lede.xian@mpsd.mpg.de}
\affiliation{Max Planck Institute for the Structure and Dynamics of Matter, Center for Free Electron Laser Science, 22761 Hamburg, Germany}

\author{Dante M. Kennes}
\affiliation{Institut f\"ur Theorie der Statistischen Physik, RWTH Aachen University and JARA-Fundamentals of Future Information Technology, 52056 Aachen, Germany}
\altaffiliation{dante.kennes@rwth-aachen.de}
\affiliation{Max Planck Institute for the Structure and Dynamics of Matter, Center for Free Electron Laser Science, 22761 Hamburg, Germany}

\author{Angel Rubio}
\affiliation{Max Planck Institute for the Structure and Dynamics of Matter, Center for Free Electron Laser Science, 22761 Hamburg, Germany}
\affiliation{Center for Computational Quantum Physics, Simons Foundation Flatiron Institute, New York, NY 10010 USA}
\altaffiliation{angel.rubio@mpsd.mpg.de}

\date{\today}

\begin{abstract}
We predict that twisted bilayers of 1T-ZrS$_2$ realize a novel and tunable platform to engineer two-dimensional topological quantum phases dominated by strong spin-orbit interactions. At small twist angles, ZrS$_2$ heterostructures give rise to an emergent and twist-controlled moir\'e Kagom\'e lattice, combining geometric frustration and strong spin-orbit coupling to give rise to a moir\'e quantum spin Hall insulator with highly controllable and nearly-dispersionless bands. We devise a generic pseudo-spin theory for group-IV transition metal dichalcogenides that relies on the two-component character of the valence band maximum of the 1T structure at $\Gamma$, and study the emergence of a robust quantum anomalous Hall phase as well as possible fractional Chern insulating states from strong Coulomb repulsion at fractional fillings of the topological moir\'e Kagom\'e bands. Our results establish group-IV transition metal dichalcogenide bilayers as a novel moir\'e platform to realize strongly-correlated topological phases in a twist-tunable setting.
\end{abstract}
\maketitle

Twisted van der Waals heterostructures have recently emerged as an intriguing and highly tunable platform to realize unconventional electronic phases in two dimensions \cite{balents20,kennes20,andrei20,andrei21} and more. Spurred by the discovery of Mott insulation and superconductivity in twisted bilayer graphene \cite{cao18a,cao18b}, remarkable progress in fabrication and twist-angle control has lead to observations of correlated insulating states or superconductivity in a variety of materials, including trilayer and double-bilayer graphene, homo- and hetero-bilayers of twisted transition metal dichalcogenides \cite{lu2019superconductors,stepanov2019interplay,BiGraphrev,shen2019observation,Kerelskye2017366118,rubioverdu2020universal,arora2020superconductivity,chen2020nature,liu2019spin,cao2019electric,he2020,Wang19TMD}, and heterostructures at a twist on hexagonal boron nitride substrates \cite{chen2018gate,chen2019signatures}.
\textbf{}
At its heart, this rich phenomenology stems from electronic interference effects due to the moir\'e superlattice, which can selectively quench kinetic energy scales to realize almost dispersionless bands, permitting a twist angle controlled realization of regimes dominated by strong electronic interactions. At the same time, the drastic reduction of kinetic energy of the low-energy moir\'e bands implies straightforward gate-tunable access to a wide range of filling fractions, permitting wide-ranging experimental access to the phase diagrams of paradigmatic models of strongly-correlated electrons \cite{kennes20}. Consequently, the putative realization of strongly-correlated electron physics in a tunable setting has garnered significant attention, resulting in growing experimental evidence for novel correlated phases, including unconventional superconductivity \cite{andrei21,lu19,cao20a}.

\begin{figure*}[t]
	\centering
	\includegraphics[width=\textwidth, trim=0 0.5cm 0 0]{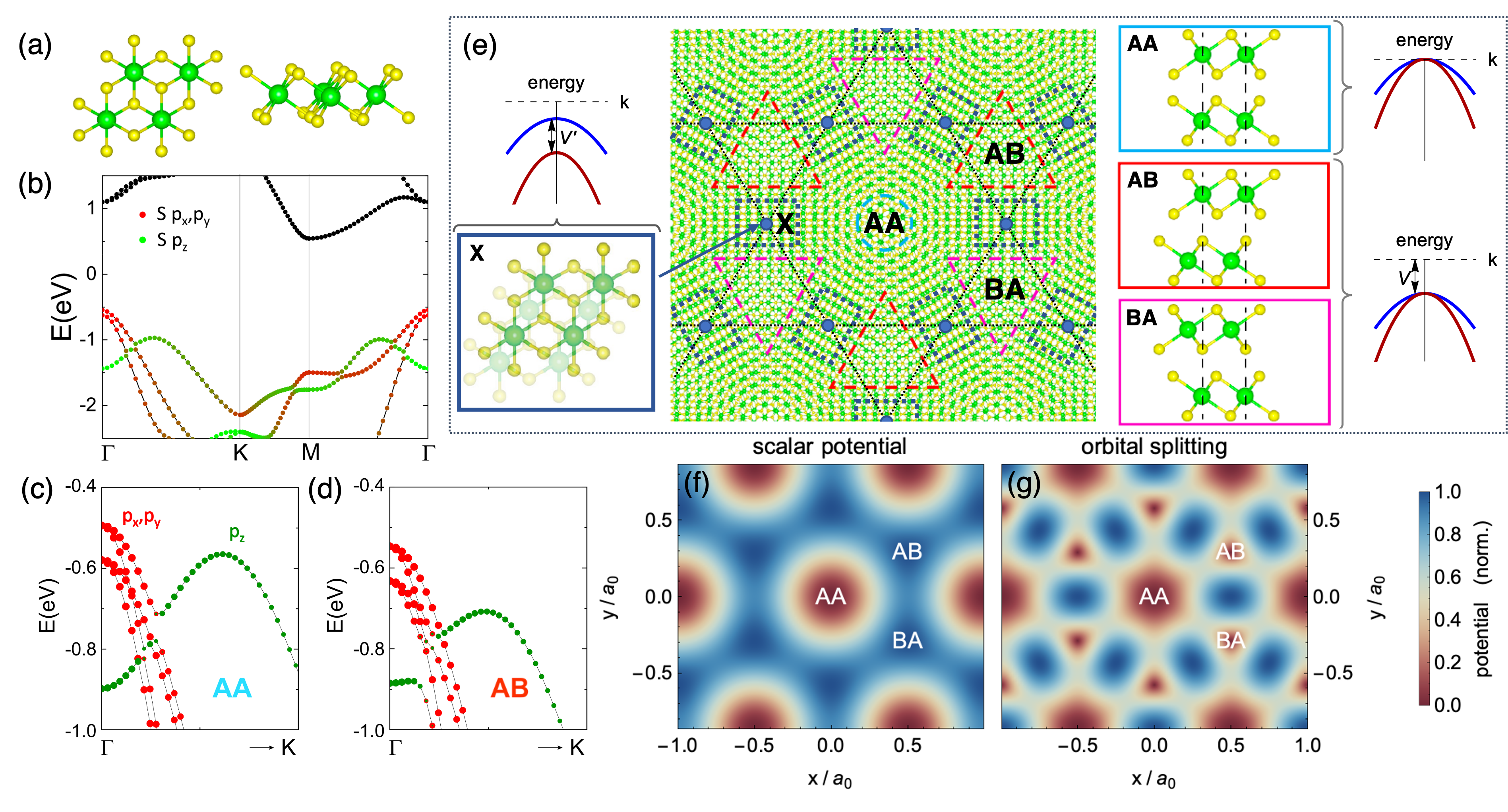}
	\caption{\captiontitle{Moir\'e patterns of twisted ZrS\textsubscript{2} bilayers.} (a) Top and side view of the 1T structure of ZrS$_2$. (b) The band structure of ZrS$_2$ monolayer hosts a valence band maximum at $\Gamma$ composed of the $p_x$, $p_y$ chalcogen orbital $J=\pm 3/2$ states, with a spin-orbit splitting of $\sim 100$ meV. (c,d) In aligned bilayers with AA or AB stacking, the valence band maximum  at $\Gamma$ (composed of of antibonding states of $p_x$, $p_y$ symmetry) is energetically separated by $\sim$ 50meV from bonding $p_x, p_y$ states and $p_z$ orbitals. Hence, antibonding $p_x$, $p_y$ states compose the top-most moir\'e valence bands at small twist angles. (e) In small-angle twisted heterostructures of ZrS$_2$, the interlayer alignment interpolates from AA to AB/BA stacking as a function of position in the moir\'e unit cell. The $p_x$, $p_y$ antibonding states remain degenerate for both AA and AB/BA stacking in the absence of spin-orbit coupling. Their overall energetic shift is encoded in (f) an effective \textit{scalar} moir\'e superlattice potential, with minima that form a moir\'e honeycomb lattice. (g) Crucially, the loss of rotational symmetry away from local AA, AB stacking further splits the $p_x$, $p_y$ states. This orbital splitting contribution to the moir\'e potential is maximal in three ``domain wall'' regions ``X'' in the moir\'e unit cell [(e)]; upon exceeding the scalar potential, electrons in the top-most moir\'e valence band form an emergent moir\'e Kagom\'e lattice [dashed lines in (e); guide to the eye].}
	\label{fig:fig1}
\end{figure*}

Notably, and despite negligible intrinsic spin-orbit coupling in graphene, these were found to include topological states of matter. Here, the realization of the interaction-induced quantum anomalous Hall effect without external magnetic fields in twisted bilayer \cite{sharpe19,serlin20,wu21} and trilayer \cite{chen20} graphene has spurred numerous proposals for more exotic fractionalized topological states of matter \cite{liu20,abouelkomsan19,repellin19,ledwith19}, which however rely on a delicate interplay of spontaneous ferromagnetic order, valley polarization and substrate engineering effects to induce the requisite non-trivial band topology. Generalizations to twisted transition-metal dichalcogenides have focused on telluride-based group VI compounds with 2H structure in the monolayer which exhibit an intrinsic quantum spin Hall effect \cite{wu18}, with the quantum anomalous Hall effect recently observed \cite{li21} and similarly expected to emerge from spontaneous valley polarization \cite{zhang21,xie21}.

Central to the present work, we demonstrate for twisted bilayer ZrS$_2$ with 1T structure that the paradigm of twist-controlled suppression of the bare kinetic energy scales can be straightforwardly extended to instead promote \textit{spin-orbit coupling} to constitute the dominant energy scale at low energies, opening up a new and exotic regime for experimental and theoretical investigation. Remarkably, we find that the two-component character of the valence band maximum in such two-dimensional group IV transition metal dichalcogenides enters in an essential manner, leading to the emergence of a clean {\it moir\'e Kagom\'e} lattice with almost dispersionless quantum spin Hall bands at small twist angles. We demonstrate that this tunable realization of a ZrS$_2$ moir\'e heterostructure with strong spin-orbit coupling and strong interactions can therefore provide a robust and novel platform to probe the profound interplay of non-trivial band topology and electronic correlations, and shed light on elusive quantum phases beyond the purview of conventional condensed matter systems.

ZrS$_2$ is an exfoliable semiconductor with 1T structure \cite{zhang2015controlled} in its ground state [Fig. \ref{fig:fig1}(a)]. In contrast to group-VI transition metal dichalcogenides such as MoS$_2$ with 2H structure, the valence band maximum in ZrS$_2$ and other group-IV transition metal dichalcogenides is located at $\Gamma$ already in the monolayer and is composed of two-fold degenerate chalcogen $p_x$, $p_y$ orbitals. Spin-orbit coupling lifts their degeneracy and introduces a $\sim 100 \textrm{meV}$ gap [Fig. \ref{fig:fig1}(b)]. This property readily carries over to aligned bilayers with symmetric AA and AB stacking configurations [Figs. \ref{fig:fig1}(c), (d)]; here, the valence band maximum at $\Gamma$ follows from antibonding combinations of the out-of-plane chalcogen $p_x$, $p_y$ orbitals. These are energetically separated from bonding combinations by $\sim 80-100$ meV [Fig. \ref{fig:fig1}(c), (d)], with a secondary local valence band maximum of $p_z$ orbitals furthermore located close to $\Gamma$ and similarly detuned by $\sim 50$ meV for AA stacking.

\begin{figure*}[t]
	\centering
	\includegraphics[width=\textwidth]{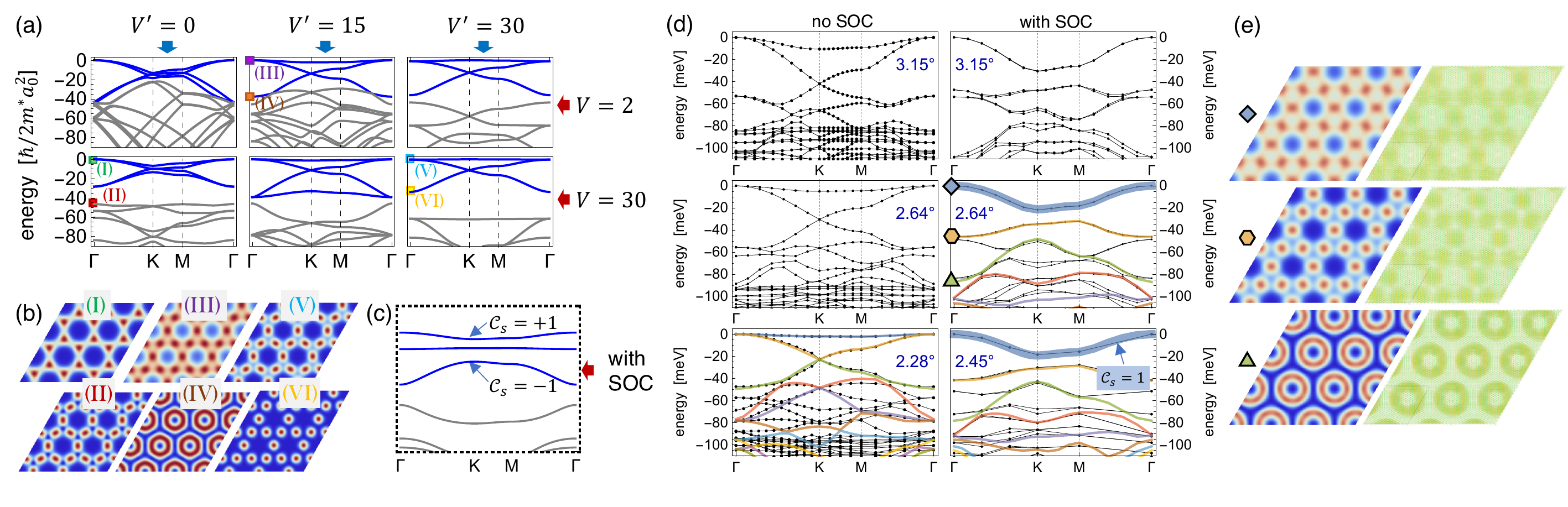}
	\caption{\captiontitle{Emergent Kagom\'e lattice and continuum theory of twisted ZrS\textsubscript{2} bilayers.} (a) Anatomy of the moir\'e band structure in the two-orbital pseudospin continuum theory, as a function of the scalar ($V$) and pseudospin ($V'$) potential, with $\eta = 0.15$ and energies normalized to $\hbar / 2m^\star a_0^2 \equiv 1$. In the absence of $V'$, a well-separated honeycomb lattice with two $p$ orbitals per sublattice emerges with increasing $V$ [bottom left], directly analogous to twisted bilayer graphene. (b) The calculated charge density of bands (I) is localized in a honeycomb pattern of AB/BA regions; however, the next lower-energy band (II) already exhibits a Kagom\'e charge density pattern. Conversely, the pseudospin moir\'e potential $V'$ favors charge patterns localized in (III) the Kagom\'e ``X'' regions as well as on rings (IV) around the Kagom\'e hexagons. As $a_0 \sim \theta^{-1}$, an energetically well-separated Kagom\'e moir\'e band structure emerges at sufficiently small twist angles [(a), right column, marked in blue]. (c) Spin-orbit interactions lift the quadratic touching of Kagom\'e bands at $\Gamma$. The resulting band structure is gapped and realizes a novel Kagom\'e moir\'e flat-band quantum spin Hall insulator with spin Chern numbers $\pm 1$ of the top and bottom Kagom\'e bands. (d) depicts the \textit{ab initio} band structure of twisted bilayers for three representative twist angles with [right column] and without [left column] spin-orbit coupling, with the top-most valence bands demonstrating the emergent Kagome lattice. Colored lines indicate the continuum model band structure, fitted to the top valence bands [see main text]. Additional deeper $p_z$ orbital valence bands appear $> 50meV$ below the band edge and are not accounted for in the continuum theory, however are progressively separated energetically from the Kagom\'e bands as the twist angle is reduced. Importantly, spin-orbit interactions split off a topological band with spin Chern number $\mathcal{C}_s = 1$ [thick blue line], as discussed in the main text. (e) Charge density distribution from our \textit{ab initio} calculations confirms the emergent Kagom\'e band structure.}
	\label{fig:fig2}
\end{figure*}

In twisted bilayers, the atomic interlayer registry interpolates continuously between local AA, AB and BA alignment as a function of position and a moir\'e pattern with three-fold rotation symmetry forms [Fig. \ref{fig:fig1}(e)]. At sufficiently small twist angles, the energetic considerations for aligned bilayers discussed above immediately suggest that the top-most (highest energy) moir\'e valence bands should be similarly composed of antibonding $p_x$, $p_y$ chalcogen orbitals. If spin-orbit coupling is neglected, these are degenerate at $\Gamma$ in both AA and AB regions [Fig. \ref{fig:fig1}(e)] by virtue of rotation symmetry. However, the valence band edge differs between the two stackings, with the smooth interpolation between local alignments in the moir\'e unit cell encoded in an effective periodic \textit{scalar} moir\'e potential $V(\r)$ [Fig. \ref{fig:fig1}(f)]. Minima of $V(\r)$ are located at the AB and BA regions and form an effective honeycomb lattice. Notably, for the purposes of capturing the highest-energy moir\'e valence bands, the model retains to an excellent approximation the full six-fold rotation and mirror symmetries of the monolayer, even though the macroscopic crystal is chiral. This situation is in principle analogous to twisted bilayer MoS\textsubscript{2} \cite{xian20,angeli20}, which hosts a series of almost dispersionless bands of Mo $d_{z^2}$ orbital character on an emergent moir\'e honeycomb lattice.

Crucially, the loss of rotational symmetry away from local AA, AB stacking lifts the orbital degeneracy between $p_x$, $p_y$ antibonding orbitals (in the absence of spin-orbit coupling), introducing a second energy scale into the problem. 
In stark contrast to 2H TMD bilayers, the two-component character of the $\Gamma$ valley states enters in an essential manner. From symmetry considerations, their orbital splitting is expected to be maximal in three ``domain wall'' regions ``X'' per moir\'e unit cell in Fig. \ref{fig:fig1}(e), in which the transition-metal atoms of both layers form ``stripes'', with the rotational symmetry of the local stacking order reduced to C$_2$. Contrary to the scalar moir\'e potential, maxima of orbital $p_x$, $p_y$ splitting hence form a Kagom\'e pattern [Fig. \ref{fig:fig1}(g)]. Remarkably, if the resulting energetic gain exceeds the scalar potential $V(\r)$, it becomes favorable for charge to migrate from the honeycomb AB/BA regions to ``X'' regions, realizing an \textit{emergent Kagom\'e lattice} of $s$-like moir\'e orbitals in a highly-tunable setting [Fig. \ref{fig:fig1}(e)].

A minimal continuum model of this scenario readily follows from the above symmetry considerations as
\begin{align}
	\Ham = \Ham_0 + \Ham_{\textrm{soc}} + \Ham_{\textrm{pot}}  \label{eq:continuum}
\end{align}
where $\Ham_0$ describes the two-fold degenerate antibonding $p_x$, $p_y$ chalcogen states
\begin{align}
	\Ham_0 &= -\frac{\hbar^2}{2m^\star} \left\{ (k_x^2 + k_y^2) \tauID + \vphantom{\frac{\eta}{2}} \eta \left[ (k_x^2 - k_y^2) \tauZ + 2 k_x k_y \tauX \right] \right\}
\end{align}
with the orbital degree of freedom represented via Pauli matrices $\hat{\boldsymbol{\tau}}$. Here, $m^\star$ denotes the effective average band mass, and $\eta = \tfrac{m_+ - m_-}{m_+ + m_-}$ parametrizes the ratio of light ($m_-$) and heavy ($m_+$) hole $p$ bands at $\Gamma$. Atomic spin-orbit interactions
\begin{align}
	\Ham_{\textrm{soc}} = \frac{\lambda_{\textrm{soc}}}{2} ~\tauY \hat{\boldsymbol{\sigma}}_z   \label{eq:SOC}
\end{align}
lift the orbital degeneracy, opening up a gap at $\Gamma$ as discussed in detail below. Here, $\hat{\boldsymbol{\sigma}}_z$ acts on spin. Central to the emergence of the Kagom\'e lattice, the moir\'e potential acts non-trivially on the orbital pseudospin, and can generically be written as a Fourier expansion
\begin{align}
	\Ham_{\textrm{pot}} = \sum_n V_n ~ f_n^{0}(\r) ~+~ \sum_n V'_n \left[ \tauX ~f_n^{(x)}(\r) + \tauZ ~f_n^{(z)}(\r) \right]
\end{align}
Here, $n$ indexes the $n$-th moir\'e Brillouin zone. $V_n$ parameterizes the Fourier modes of the scalar potential in direct analogy to twisted WS$_2$ \cite{angeli20}, with $f_n^{(0)}(\r) = \cos(\mathbf{b}_{n,1} \r) + \cos(\mathbf{b}_{n,2} \r) + \cos(\mathbf{b}_{n,3} \r)$ chosen to retain the full six-fold rotation symmetry and $\mathbf{b}_{n,i}$ describing the three reciprocal lattice vectors $i=1,2,3$ (related via $C_3$ rotations) to the $n$-th Brillouin zone.


\begin{figure*}[t]
	\centering
	\includegraphics[width=0.9\textwidth]{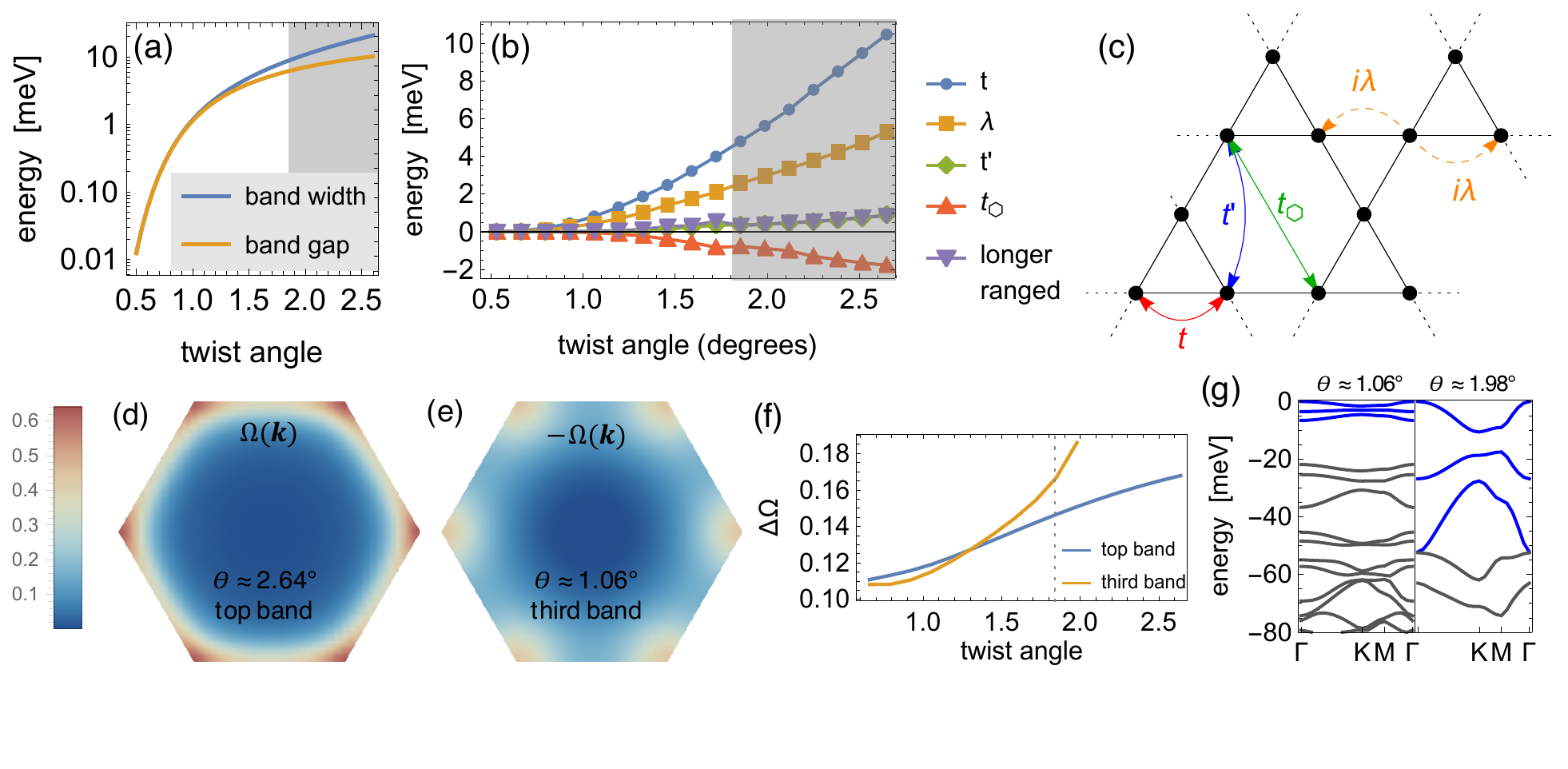}
	\caption{\captiontitle{Twist angle dependence of the Kagom\'e Moir\'e lattice.} (a) Band width and energetic separation of the top-most Moir\'e valence band, extrapolated from the continuum theory as a function of twist angle. (b) Kagom\'e tight-binding parameterization of the top three Moir\'e bands, with real and imaginary hoppings depicted schematically in (c). Shaded regions denote larger twist angles for which the third Kagom\'e band does not remain well-isolated from lower-lying states; tight-binding parameterization in this region fits only the top two bands. (d) Berry curvature at $2.64^{\circ}$ for the first Moir\'e Kagom\'e valence band, as well as (e) for the third band at $1.06^{\circ}$. (f) Berry curvature fluctuations $\Delta\Omega$ are suppressed as the twist angle is reduced, approaching a Moir\'e realization of a Landau level. (g) Band structure of the continuum model at $\theta \approx 1.06^{\circ}, 1.98^{\circ}$.}
	\label{fig:tightbinding}
\end{figure*}

The pseudospin $\tauX$, $\tauZ$ contributions to the potential are related in the presence of (approximate) mirror symmetry, with $f_n^{(x)}(\r) = -\frac{\sqrt{3}}{2} \cos(\mathbf{b}_{n,1} \r) + \frac{\sqrt{3}}{2} \cos(\mathbf{b}_{n,3} \r)$ and $f_n^{(z)}(\r) = \frac{1}{2} \cos(\mathbf{b}_{n,1} \r) - \cos(\mathbf{b}_{n,2} \r) + \frac{1}{2} \cos(\mathbf{b}_{n,3} \r)$. The salient physics is encoded already in the lowest harmonic -- with $\mathbf{b}_{1,1} = [2\pi, -2\pi/\sqrt{3} ] / a_0$, $\mathbf{b}_{1,2} = [0, 4\pi/\sqrt{3}]/a_0$, $\mathbf{b}_{1,3} = [2\pi, 2\pi/\sqrt{3}] / a_0$ and $a_0$ the moir\'e lattice length, the scalar potential hosts two minima in the AB and BA regions at $\mathbf{r} = [1/2, \pm 1 / (2\sqrt{3}) ]$. Conversely, the pseudospin potential that determines the splitting of $p_x$, $p_y$ orbitals has three maxima in the Kagom\'e X regions at $\mathbf{r} = [1/2, 0] a_0 ,~ [1/4, \sqrt{3}/4] a_0 ,~ [3/4, \sqrt{3}/4] a_0$. 
Up to an overall energy scale, a minimal continuum model that includes only the first harmonic will therefore depend on just three dimensionless parameters $\eta$, $2 V m^\star a_0^2 / \hbar^2$, $2 V' m^\star a_0^2 / \hbar^2$, where $a_0 \propto \theta^{-1}$ scales with the twist angle. Local lattice relaxation effects are encoded in the higher harmonics of the potential; as the local stacking of AB and BA regions arise energetic favorable, lattice relaxation results in large domains with almost uniform AB or BA stacking [see domains highlighted with red and purple dashed lines in Fig.~1 (e)] with the salient stacking variation in the X regions. These parameters can be obtained fitting the band structure obtained from DFT calculations as described below.

Fig. \ref{fig:fig2}(a) depicts the structure of the resulting moir\'e bands without spin-orbit coupling, as a function of scalar $V \equiv V_1 $and pseudospin $V' \equiv V'_1$ potentials. For $V' = 0$, the scalar potential $V$ localizes the hole charge density on a honeycomb lattice of AB/BA regions [Fig. \ref{fig:fig2}(a), left column; Fig. \ref{fig:fig2}(b) (I)], and an energetically well-separated set of honeycomb bands with Dirac points at $\mathbf{K}$, $\mathbf{K}'$ emerges at the top of the valence band. These retain a two-fold orbital $p_x$, $p_y$ character, with the degeneracy of the bands weakly broken due to orbital anisotropy $\eta \neq 0$. This directly mirrors the low-energy band structure of twisted bilayer graphene, however with the two-orbital structure resulting from the  $p_x$, $p_y$ degeneracy of the constituent states at $\Gamma$ as opposed to a valley degeneracy.

However, already the next lower in energy (fifth) moir\'e valence band reveals upon closer inspection a charge density distribution with a Kagom\'e pattern [Fig. \ref{fig:fig2}(b), pattern (II)], localized in the ``X'' regions of the moir\'e unit cell [Fig. \ref{fig:fig1}(e)]. These states gain energy from a finite pseudospin potential $V'$, which lifts them to higher energies: Beyond a critical $V'$, the fifth ``Kagom\'e'' band and the bottom $p_x$, $p_y$ honeycomb bands invert their energetic ordering at $\Gamma$. Consequently, the charge density distribution of the top $p_x$, $p_y$ bands shifts from AB/BA honeycomb regions to ``X'' Kagom\'e sites [Fig. \ref{fig:fig2}(b), pattern (III)]. If the moir\'e potentials are sufficiently weak, the three resulting bands that constitute the emergent moir\'e Kagom\'e lattice couple to a fourth moir\'e orbital centered on the hexagons of the lattice, with a charge density distribution that forms a ring around the AA regions of the moir\'e unit cell [Fig. \ref{fig:fig2}(b), pattern (IV)]. As the twist angle is further reduced, an energetically well-separated set of three Kagom\'e lattice bands emerges as the top-most set of moir\'e valence states [Fig. \ref{fig:fig2}(b), patterns (V), (VI)].


The above behavior closely matches the results from large-scale \textit{ab initio} calculations of the twisted moir\'e supercell, depicted in Fig. \ref{fig:fig2}(d), left column, for three representative twist angles [also see Supplementary Information]. As the angle is reduced, a set of bands with a Kagom\'e charge distribution at $\Gamma$ splits off progressively from deeper valence bands. For the larger twist angles $\geq 2.28^{\circ}$ that are still within computational reach for density functional calculations, this energetic separation is not yet sufficient to completely separate the Kagom\'e bands of chalcogen antibonding $p_x$, $p_y$ character from states with $p_z$ or bonding $p_x$, $p_y$ character ($<50 {\rm meV}$ below the band edge), not included in the continuum theory. Nevertheless, the top-most Kagom\'e bands of interest are already well-captured via the continuum model for the smallest twist angle [Fig. \ref{fig:fig2}(d), bottom-left] upon accounting only for the lowest harmonic of the moir\'e potential.

Crucially, the inclusion of spin-orbit coupling [Eq. (\ref{eq:SOC})] now opens up a gap at the Kagom\'e Dirac points and lifts the quadratic band touching degeneracy at $\Gamma$ [Fig. \ref{fig:fig2}(c)], reflected in \textit{ab initio} simulations with spin-orbit interactions [Fig. \ref{fig:fig2}(d), right column]. As the top-most valence states originate from $p_x$, $p_y$ orbitals at small twist angles, spin-flip spin-orbit interactions are negligible and spin-$z$ remains a good quantum number. Remarkably, this results in three almost dispersionless moir\'e bands that realize a novel Kagom\'e topological quantum spin Hall insulator with spin Chern numbers $\mathcal{C}_s = \pm 1$ for the first and third flat band [Fig. \ref{fig:fig2}(c), (d)]. In marked contrast to conventional topological materials however, while superlattice interference quenches the kinetic energy scales, spin-orbit coupling $\lambda_{\textrm{soc}}$ enters as a bare atomic scale and hence becomes the \textit{dominant energy scale} that governs the low-energy physics of the moir\'e valence bands in ZrS$_2$. This highly-tunable materials realization of an ``ultra-strong'' spin-orbit interaction regime in a moir\'e heterostructure constitutes a central result of this paper.

To model the emergent top-most flat topological moir\'e band in twisted ZrS$_2$, we proceed with a fit of the pseudospin continuum theory [Eq. (\ref{eq:continuum})] to the spin-orbit-coupled \textit{ab initio} band structure for $\theta = 2.64^{\circ}$ [Fig. \ref{fig:fig2}(d), middle-right panel]. As the minimal model of Eq. (\ref{eq:continuum}) does not account for bonding $p_x, p_y$ or $p_z$ states, the third-highest \textit{ab initio} valence band ($-50{\rm meV}$ below the valence band edge) is composed primarily of bonding $p_x, p_y$ and $p_z$ orbitals and is excluded from the fit. We note that this band separates energetically from the three Kagom\'e moir\'e bands at lower twist angles. We obtain excellent agreement for the top two bands of $p_x$, $p_y$ antibonding character using $\eta = 0.33$, $m^\star = 0.27m_0$, $\lambda_{\textrm{soc}} = 57 {\rm meV}$, $V_1 = 5.5{\rm meV}$, $V'_1 = -9.3{\rm meV}$, $V_2 = 11.5{\rm meV}$, $V'_2 = -5.1{\rm meV}$. Scaling with twist angle similarly matches the \textit{ab initio} band structure at $2.45^{\circ}$ [Fig. \ref{fig:fig2}(d), bottom-right panel]. As expected, the top-most band is topologically non-trivial with spin Chern number $\mathcal{C}_s = \pm 1$. Fig. \ref{fig:fig2}(e) compares the corresponding charge density distributions at $\Gamma$ for \textit{ab initio} and continuum model calculations; both exhibit comparable Kagom\'e patterns as well as a competing band at lower energies with a ring-shaped charge pattern around the $AA$ region, which similarly becomes energetically separated from Kagom\'e bands at lower twist angles [Fig. \ref{fig:fig2}(a)].

\begin{figure*}[t]
	\centering
	\includegraphics[width=\textwidth]{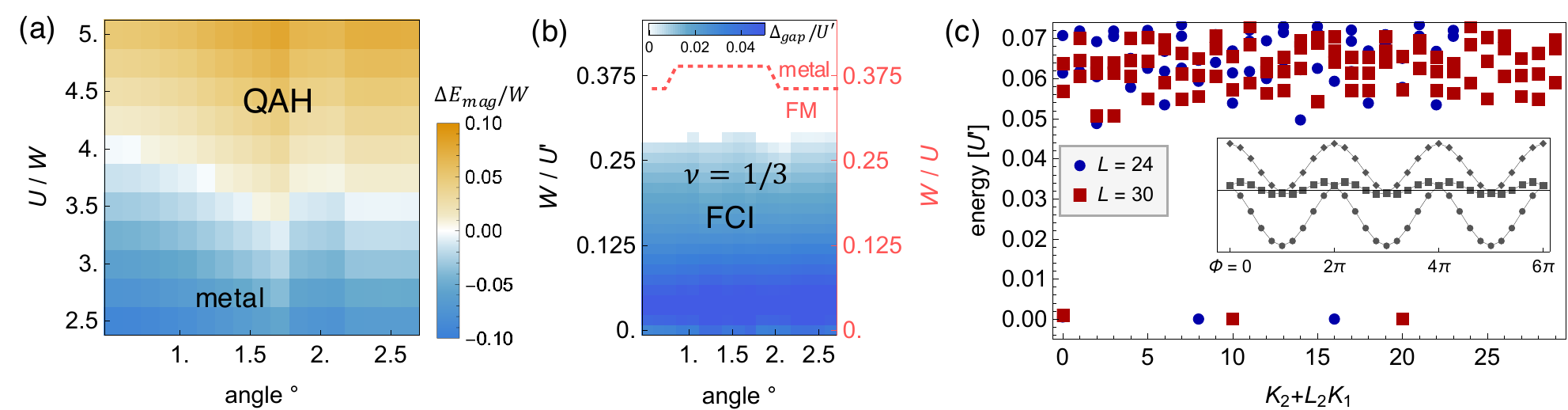}
	\caption{\captiontitle{Quantum anomalous Hall effect and fractional Chern insulators.} (a) At half filling, the first Moir\'e band displays a ferromagnetic instability for a local Hubbard repulsion $U$ on the order of $\sim 4W$ the electronic bandwidth, depicted via the energy density gain from spontaneous spin polarization. Spin polarized Moir\'e quantum spin Hall bands realize a correlated quantum anomalous Hall phase. (b) At $1/6$ hole doping, ferromagnetism (FM) persists for finite $U$ [right axis] over a wide range of twist angles. Finite longer-ranged Coulomb repulsion $U'$ [left axis] between nearest-neighbor Kagom\'e sites drives this spin-polarized metal into a $\nu = 1/3$ Laughlin phase, realizing a robust fractional Chern insulator, depicted in terms of the many-body excitation gap. (c) Fingerprint of the fractional Chern insulator in the low-energy spectrum for $L=24$ ($L=30$) unit cells with periodic boundary conditions and $N=8$ ($N=10$) fermions, depicted as a function of linearized total momentum $(K_1,K_2)$. Inset depicts the characteristic spectral flow under insertion of a magnetic flux through the torus.}  \label{fig:interactions}
\end{figure*}

A key advantage of the continuum theory is the possibility to study the behavior at small twist angles in a computationally feasible manner. Fig. \ref{fig:tightbinding}(a) depicts the band width of the top-most topological moir\'e Kagom\'e band, as well as the single-particle gap to the next deeper valence band, as a function of twist angle $a_0 \sim \theta^{-1}$. The band width of the top-most topological band decreases exponentially with twist angle, whereas the ratio between band width and band gap saturates below $\approx 2^{\circ}$ and approaches one. Below this twist angle, the three Kagom\'e bands become fully isolated in energy from deeper valence states [Fig. \ref{fig:tightbinding}(g)]. This immediately suggests a fruitful tight-binding parameterization at ultra-small angles, presuming that local lattice relaxation effects remain manageable. Results are shown in Fig. \ref{fig:tightbinding}(b) for a tight-binding model depicted schematically in (c), but including up to 8th-neighbor hopping to ensure a good fit over all angles [see Supplementary Information]. For small angles $\ll 2^{\circ}$, the top three bands become well-captured by a nearest-neighbor Kagom\'e tight-binding model with imaginary hoppings. Third-neighbor hopping $t_{\varhexagon}$ through the hexagons are leading corrections to this model and follow from the elliptical shapes of the charge density distribution at the Kagom\'e ``X'' sites.

The sizable imaginary nearest-neighbor hopping [Fig. \ref{fig:tightbinding}(b)] is a direct consequence of the strong spin-orbit coupling limit and can be interpreted as a finite effective staggered magnetic flux through the elementary triangles of the Kagom\'e lattice. It lifts the quadratic touching of flat and dispersive Kagom\'e bands and opens up a gap at the Dirac points, realizing a time-reversal-invariant version of a parent model for fractional Chern insulators \cite{tang10,wu11}. Here, uniformity of the Berry curvature is a key figure of merit \cite{parameswaran11} and determines, jointly with the ideal droplet condition for the Fubini-Study metric \cite{claassen15,jackson14,lee17}, the propensity for flat Chern bands to host fractional quantum Hall phases. Figs. \ref{fig:tightbinding}(d), (e) depict the Berry curvature for the top-most and third topological Kagom\'e moir\'e band, for two representative twist angles, with Berry curvature fluctuations $(\Delta\Omega)^2 = \int_{\textrm{BZ}} [ \Omega(\k) - \sqrt{3} \mathcal{C} / 8\pi^2 ]^2 d\k $ quantified in (f), where $\mathcal{C} = \pm 1$ is the Chern number. The Berry curvature flattens monotonically as the twist angle is reduced, with fluctuations substantially suppressed for the third Kagom\'e valence band at small angles.

The tunable realization of isolated topological flat bands in twisted ZrS$_2$ is an ideal starting point for the stabilization of a host of correlated topological states of matter, ranging from the quantum anomalous Hall effect to elusive fractional Chern insulator and fractional topological insulator phases. To investigate the role of electronic correlations, we augment the effective tight-binding description [Fig. \ref{fig:tightbinding}] via a screened Coulomb repulsion, constrained for simplicity to a local Hubbard ($U \sum_i \ND{i\uparrow} \ND{i\downarrow}$) and nearest-neighbor density ($(U'/2) \sum_{\left<ij\right>\sigma\sigma'} \ND{i\sigma} \ND{j\sigma'}$) interaction. Suppose first that the top-most topological moir\'e Kagom\'e band is tuned to half filling via electrostatic gating. A non-trivial spin Chern number precludes a straightforward Wannier tight-binding representation of the band. Instead, as deeper fully-filled valence bands are energetically separated, the low-energy behavior can be captured via projecting Coulomb interactions $U, U'$ to the half-filled flat topological band, in direct analogy to lowest Landau level projections for the fractional quantum Hall effect. The resulting interacting problem is governed by an effective Hamiltonian
\begin{align}
	\Ham_{\textrm{eff}} &= \sum_{\k\sigma} \E_{\k} \CD{\k\sigma} \C{\k\sigma} + \frac{1}{L} \sum_{\substack{\k\k'\q \\ \sigma\sigma'}} V_{\k\k'\q}^{\sigma\sigma'} ~\C{\k,\sigma} \C{\k',\sigma'} \CD{\k'-\q,\sigma'} \CD{\k+\q,\sigma}  \label{eq:HamProjected}
\end{align}
where $\CD{\k\sigma}, \C{\k\sigma}$ create/annihilate electrons in the flat band, $\E_{\k}$ denotes the residual band dispersion, $L$ is the system size, and
\begin{align}
    V_{\k\k'\q}^{\sigma\sigma'} = \frac{1}{2} \sum_{\alpha\alpha'} v_{\alpha\alpha'}(\q)~ u_{\k}^{(\alpha\sigma)} u_{\k'}^{(\alpha'\sigma')} \left[ u_{\k'-\q}^{(\alpha'\sigma')} u_{\k+\q}^{(\alpha\sigma)} \right]^\star
\end{align}
is the Coulomb repulsion projected to the Bloch states $u_{\k}^{(\alpha\sigma)}$ of the top-most band, with
\begin{align}
    v(\q) = \left[\begin{array}{ccc} U & U' \cos(\tfrac{\k\mathbf{a}_1}{2}) & U' \cos(\tfrac{\k\mathbf{a}_2}{2}) \\
    U' \cos(\tfrac{\k\mathbf{a}_1}{2}) & U & U' \cos(\tfrac{\k\mathbf{a}_3}{2}) \\
    U' \cos(\tfrac{\k\mathbf{a}_2}{2}) & U' \cos(\tfrac{\k\mathbf{a}_3}{2}) & U \end{array}\right]    
\end{align}
Here, momenta $\k,\k',\q$ are defined in the moir\'e Brillouin zone, $\mathbf{a}_i$ denote the Moir\'e lattice vectors, and $\alpha,\sigma$ denote the sublattice and spin degrees of freedom. 

Since a sufficiently short-ranged interaction $U > U'$ mainly imparts a local energetic penalty for electron pairs of opposite spin occupying the same Kagom\'e ``X'' sites, a flat-band ferromagnetic instability generically ensues \cite{mielke91} at half filling of the top-most quantum spin Hall band, in direct analogy to quantum Hall ferromagnetism \cite{sondhi93}. The resulting spontaneous spin-polarized state is gapped and aligned in the $z$ direction -- it exhibits a quantum anomalous Hall effect by virtue of filling a quantum spin Hall band for one spin component, and becomes an exact zero-energy ground state in the absence of dispersion \cite{neupert11}. Fig. \ref{fig:interactions}(a) depicts the corresponding phase diagram as a function of twist angle and interaction strength $U$ vs band width $W$ of the top-most band, evaluated from exact diagonalization of Eq. (\ref{eq:HamProjected}) on a $4 \times 4$ unit cell cluster.  A robust quantum anomalous Hall state emerges for interactions on the order of four times the moir\'e band width and remains robust over a wide range of twist angles. Notably, the underlying mechanism is distinct from the observed quantum anomalous Hall effect in twisted bilayer graphene, relying instead on the \textit{intrinsic} topologically non-trivial moir\'e band structure due to strong spin-orbit coupling and obviating the necessity for concurrent valley polarization and substrate effects.

Persistence of the ferromagnetic instability for fractional fillings of the Kagome\'e flat bands naturally suggests the possibility to stabilize a variety of Abelian and non-Abelian fractional quantum Hall states in the absence of external magnetic fields \cite{bergholtz13}, by analogy to a fractionally-filled Landau level. To this end, we focus on the Laughlin $\nu = 1/3$ state at $1/6$ hole doping, and study the interacting problem at small twist angles in exact diagonalization. Analogous to the half-filled case, electrons in the almost-flat band can avoid local Coulomb repulsion $U$ via spontaneous spin polarization, yielding a robust ferromagnetic instability as a function of $U$ [Fig. \ref{fig:interactions}(b), right axis, dashed line] over all investigated twist angles. However, spontaneous spin polarization due to $U$ now leaves a single Chern band at $1/3$ hole doping, with the resulting electronic phase governed by longer-ranged Coulomb interactions $U'$. To study the propensity to realize a Laughlin state, we numerically investigate the resulting phase diagram as function of bandwidth $W / U'$ [Fig. \ref{fig:interactions}(b), left axis]. For $W = 0$, corresponding to the Landau level limit of a perfectly-flat Chern band, exact diagonalization calculations for $6\times 5$ unit cells indicate the robust stabilization of a fractional Chern insulator. This phase is characterized by a three-fold ground state degeneracy for periodic boundary conditions [Fig. \ref{fig:interactions}(c)] with a gap to well-separated many-body excitations which persists as a function of system size. These ground states lie in three total momentum sectors that match the generalized Pauli principle for FCIs \cite{regnault11}, flow into each other upon adiabatic insertion of a magnetic flux through handles of the torus (periodic boundary conditions) and remain energetically separated from excitations, confirming the $\nu = 1/3$ FCI \cite{regnault11,wu11}. These conclusions remain largely independent of the twist angle, and the FCI persists upon inclusion of finite band dispersion $W$ until the many-body excitation gap closes for $W / U' \sim 0.3$ [Fig. \ref{fig:interactions}(b), false color].

Having established a robust correlated quantum anomalous Hall phase at half filling and evidence for a $\nu = 1/3$ fractional Chern insulator at one-sixth hole doping, an interesting follow-up question concerns the role of proximal deeper moir\'e valence bands, beyond the single-band approximation. For interactions that exceed the single-particle gap to other bands but remain smaller than the \textit{overall} band width of the three Kagom\'e bands, the robustness of fractional Chern insulator phases has been well-documented \cite{kourtis14}, in direct analogy to Landau level mixing in the conventional quantum Hall effect. A more substantial challenge however stems from details of possible longer-ranged electron interactions and exchange processes, which could serve to either enhance or suppress the stability of the fractionalized phases at different filling fractions. These processes sensitively depend on the screening environment and gating \cite{throckmorton11}, and microscopic calculations present a substantial methodological obstacle for twisted materials \cite{pizarro19,vanhala2020}. Conversely, for sufficiently small twist angles, if the Coulomb repulsion exceeds the overall band width of the three Kagom\'e bands, sufficient screening could serve to form a local moment at overall half filling $\nu = 3/2$. Such a Kagom\'e Mott insulator would constitute a Moir\'e realization of a paradigmatic frustrated magnetic model, which has been under intense scrutiny for the potential to host an elusive quantum spin liquid phase.

Beyond the (fractional) quantum anomalous Hall effect, the realization of flat-band quantum spin Hall insulators further opens up the possibility to realize a myriad of unconventional ordered states of matter with non-trivial topology, including time-reversal invariant fractionalized phases or topological superconductors. Consequently, twisted ZrS$_2$ bilayers constitute a promising and tunable materials platform for such investigations, granting access to a novel and exotic regime of ultra-strong spin-orbit coupling that is not readily realizable in conventional crystalline solid-state systems. More broadly, a natural question concerns the extension of similar ideas of pseudospin potential engineering and strong spin-orbit coupling to other transition-metal dichalcogenide heterostructures such as TiS$_2$ and HfS$_2$ with multi-component character of the valence band edge. At the same time, the emergence of a moir\'e Kagom\'e lattice from the fortuitous but robust interplay of geometry and interlayer coupling at small twist angles opens up a new pathway towards a moir\'e realization of magnetic phases in a paradigmatic frustrated system.



\subsection*{Acknowledgments}
This work is supported by the European Research Council (ERC-2015-AdG-694097), Grupos Consolidados (IT1249-19),  and SFB925. MC is supported by a startup grant from the University of Pennsylvania. AR is supported by the Flatiron Institute, a division of the Simons Foundation. We  acknowledge  funding by the Deutsche Forschungsgemeinschaft (DFG, German Research Foundation) under RTG 1995, within the Priority Program SPP 2244 ``2DMP'', under Germany's Excellence Strategy - Cluster  of  Excellence and Advanced Imaging of Matter (AIM) EXC 2056 - 390715994 and RTG 2247. LX acknowledges the support from Distinguished Junior Fellowship program by the South Bay Interdisciplinary Science Center in the Songshan Lake Materials Laboratory and the Key-Area Research and Development Program of Guangdong Province of China (Grants No.2020B0101340001). We acknowledge computational resources provided by the Simons Foundation Flatiron Institute, the Max Planck Computing and Data Facility and the Platform for Data-Driven Computational Materials Discovery of the Songshan Lake laboratory. This work was supported by the Max Planck-New York City Center for Nonequilibrium Quantum Phenomena.

\bibliography{ZrS2}

\end{document}

%% file: macros.tex
\newcommand{\Ham}{\hat{H}}

\newcommand{\OPc}[2]{\hat{#1}_{#2}^{\dag}}
\newcommand{\OP}[2]{\hat{#1}_{#2}^{\vphantom{\dag}}}
\newcommand{\CD}[1]{\OPc{c}{#1}}
\newcommand{\C}[1]{\OP{c}{#1}}

\newcommand{\E}{\epsilon}

%% file: ZrS2.bbl
\begin{thebibliography}{52}%
\makeatletter
\providecommand \@ifxundefined [1]{%
 \@ifx{#1\undefined}
}%
\providecommand \@ifnum [1]{%
 \ifnum #1\expandafter \@firstoftwo
 \else \expandafter \@secondoftwo
 \fi
}%
\providecommand \@ifx [1]{%
 \ifx #1\expandafter \@firstoftwo
 \else \expandafter \@secondoftwo
 \fi
}%
\providecommand \natexlab [1]{#1}%
\providecommand \enquote  [1]{``#1''}%
\providecommand \bibnamefont  [1]{#1}%
\providecommand \bibfnamefont [1]{#1}%
\providecommand \citenamefont [1]{#1}%
\providecommand \href@noop [0]{\@secondoftwo}%
\providecommand \href [0]{\begingroup \@sanitize@url \@href}%
\providecommand \@href[1]{\@@startlink{#1}\@@href}%
\providecommand \@@href[1]{\endgroup#1\@@endlink}%
\providecommand \@sanitize@url [0]{\catcode `\\12\catcode `\$12\catcode
  `\&12\catcode `\#12\catcode `\^12\catcode `\_12\catcode `\%12\relax}%
\providecommand \@@startlink[1]{}%
\providecommand \@@endlink[0]{}%
\providecommand \url  [0]{\begingroup\@sanitize@url \@url }%
\providecommand \@url [1]{\endgroup\@href {#1}{\urlprefix }}%
\providecommand \urlprefix  [0]{URL }%
\providecommand \Eprint [0]{\href }%
\providecommand \doibase [0]{http://dx.doi.org/}%
\providecommand \selectlanguage [0]{\@gobble}%
\providecommand \bibinfo  [0]{\@secondoftwo}%
\providecommand \bibfield  [0]{\@secondoftwo}%
\providecommand \translation [1]{[#1]}%
\providecommand \BibitemOpen [0]{}%
\providecommand \bibitemStop [0]{}%
\providecommand \bibitemNoStop [0]{.\EOS\space}%
\providecommand \EOS [0]{\spacefactor3000\relax}%
\providecommand \BibitemShut  [1]{\csname bibitem#1\endcsname}%
\let\auto@bib@innerbib\@empty
\bibitem [{\citenamefont {Balents}\ \emph {et~al.}(2020)\citenamefont
  {Balents}, \citenamefont {Dean}, \citenamefont {Efetov},\ and\ \citenamefont
  {Young}}]{balents20}%
  \BibitemOpen
  \bibfield  {author} {\bibinfo {author} {\bibfnamefont {L.}~\bibnamefont
  {Balents}}, \bibinfo {author} {\bibfnamefont {C.~R.}\ \bibnamefont {Dean}},
  \bibinfo {author} {\bibfnamefont {D.~K.}\ \bibnamefont {Efetov}}, \ and\
  \bibinfo {author} {\bibfnamefont {A.~F.}\ \bibnamefont {Young}},\ }\href@noop
  {} {\bibfield  {journal} {\bibinfo  {journal} {Nature Phys.}\ }\textbf
  {\bibinfo {volume} {16}},\ \bibinfo {pages} {725} (\bibinfo {year}
  {2020})}\BibitemShut {NoStop}%
\bibitem [{\citenamefont {Kennes}\ \emph {et~al.}(2021)\citenamefont {Kennes},
  \citenamefont {Claassen}, \citenamefont {Xian}, \citenamefont {Georges},
  \citenamefont {Millis}, \citenamefont {Hone}, \citenamefont {Dean},
  \citenamefont {Basov}, \citenamefont {Pasupathy},\ and\ \citenamefont
  {Rubio}}]{kennes20}%
  \BibitemOpen
  \bibfield  {author} {\bibinfo {author} {\bibfnamefont {D.~M.}\ \bibnamefont
  {Kennes}}, \bibinfo {author} {\bibfnamefont {M.}~\bibnamefont {Claassen}},
  \bibinfo {author} {\bibfnamefont {L.}~\bibnamefont {Xian}}, \bibinfo {author}
  {\bibfnamefont {A.}~\bibnamefont {Georges}}, \bibinfo {author} {\bibfnamefont
  {A.~J.}\ \bibnamefont {Millis}}, \bibinfo {author} {\bibfnamefont
  {J.}~\bibnamefont {Hone}}, \bibinfo {author} {\bibfnamefont {C.~R.}\
  \bibnamefont {Dean}}, \bibinfo {author} {\bibfnamefont {D.~N.}\ \bibnamefont
  {Basov}}, \bibinfo {author} {\bibfnamefont {A.}~\bibnamefont {Pasupathy}}, \
  and\ \bibinfo {author} {\bibfnamefont {A.}~\bibnamefont {Rubio}},\
  }\href@noop {} {\bibfield  {journal} {\bibinfo  {journal} {Nature Phys.}\
  }\textbf {\bibinfo {volume} {17}},\ \bibinfo {pages} {155} (\bibinfo {year}
  {2021})}\BibitemShut {NoStop}%
\bibitem [{\citenamefont {Andrei}\ and\ \citenamefont
  {MacDonald}(2020)}]{andrei20}%
  \BibitemOpen
  \bibfield  {author} {\bibinfo {author} {\bibfnamefont {E.~Y.}\ \bibnamefont
  {Andrei}}\ and\ \bibinfo {author} {\bibfnamefont {A.~H.}\ \bibnamefont
  {MacDonald}},\ }\href@noop {} {\bibfield  {journal} {\bibinfo  {journal}
  {Nature Mat.}\ }\textbf {\bibinfo {volume} {19}},\ \bibinfo {pages} {1265}
  (\bibinfo {year} {2020})}\BibitemShut {NoStop}%
\bibitem [{\citenamefont {Andrei}\ \emph {et~al.}(2021)\citenamefont {Andrei},
  \citenamefont {Efetov}, \citenamefont {Jarillo-Herrero}, \citenamefont
  {MacDonald}, \citenamefont {Mak}, \citenamefont {Senthil}, \citenamefont
  {Tutuc}, \citenamefont {Yazdani},\ and\ \citenamefont {Young}}]{andrei21}%
  \BibitemOpen
  \bibfield  {author} {\bibinfo {author} {\bibfnamefont {E.~Y.}\ \bibnamefont
  {Andrei}}, \bibinfo {author} {\bibfnamefont {D.~K.}\ \bibnamefont {Efetov}},
  \bibinfo {author} {\bibfnamefont {P.}~\bibnamefont {Jarillo-Herrero}},
  \bibinfo {author} {\bibfnamefont {A.~H.}\ \bibnamefont {MacDonald}}, \bibinfo
  {author} {\bibfnamefont {K.~F.}\ \bibnamefont {Mak}}, \bibinfo {author}
  {\bibfnamefont {T.}~\bibnamefont {Senthil}}, \bibinfo {author} {\bibfnamefont
  {E.}~\bibnamefont {Tutuc}}, \bibinfo {author} {\bibfnamefont
  {A.}~\bibnamefont {Yazdani}}, \ and\ \bibinfo {author} {\bibfnamefont
  {A.~F.}\ \bibnamefont {Young}},\ }\href@noop {} {\bibfield  {journal}
  {\bibinfo  {journal} {Nature Rev. Mater.}\ }\textbf {\bibinfo {volume} {6}},\
  \bibinfo {pages} {201} (\bibinfo {year} {2021})}\BibitemShut {NoStop}%
\bibitem [{\citenamefont {Cao}\ \emph {et~al.}(2018{\natexlab{a}})\citenamefont
  {Cao}, \citenamefont {Fatemi}, \citenamefont {Demir}, \citenamefont {Fang},
  \citenamefont {Tomarken}, \citenamefont {Luo}, \citenamefont
  {Sanchez-Yamagishi}, \citenamefont {Watanabe}, \citenamefont {Taniguchi},
  \citenamefont {Kaxiras}, \citenamefont {Ashoori},\ and\ \citenamefont
  {Jarillo-Herrero}}]{cao18a}%
  \BibitemOpen
  \bibfield  {author} {\bibinfo {author} {\bibfnamefont {Y.}~\bibnamefont
  {Cao}}, \bibinfo {author} {\bibfnamefont {V.}~\bibnamefont {Fatemi}},
  \bibinfo {author} {\bibfnamefont {A.}~\bibnamefont {Demir}}, \bibinfo
  {author} {\bibfnamefont {S.}~\bibnamefont {Fang}}, \bibinfo {author}
  {\bibfnamefont {S.~L.}\ \bibnamefont {Tomarken}}, \bibinfo {author}
  {\bibfnamefont {J.~Y.}\ \bibnamefont {Luo}}, \bibinfo {author} {\bibfnamefont
  {J.~D.}\ \bibnamefont {Sanchez-Yamagishi}}, \bibinfo {author} {\bibfnamefont
  {K.}~\bibnamefont {Watanabe}}, \bibinfo {author} {\bibfnamefont
  {T.}~\bibnamefont {Taniguchi}}, \bibinfo {author} {\bibfnamefont
  {E.}~\bibnamefont {Kaxiras}}, \bibinfo {author} {\bibfnamefont {R.~C.}\
  \bibnamefont {Ashoori}}, \ and\ \bibinfo {author} {\bibfnamefont
  {P.}~\bibnamefont {Jarillo-Herrero}},\ }\href@noop {} {\bibfield  {journal}
  {\bibinfo  {journal} {Nature}\ }\textbf {\bibinfo {volume} {556}},\ \bibinfo
  {pages} {80} (\bibinfo {year} {2018}{\natexlab{a}})}\BibitemShut {NoStop}%
\bibitem [{\citenamefont {Cao}\ \emph {et~al.}(2018{\natexlab{b}})\citenamefont
  {Cao}, \citenamefont {Fatemi}, \citenamefont {Fang}, \citenamefont
  {Watanabe}, \citenamefont {Taniguchi}, \citenamefont {Kaxiras},\ and\
  \citenamefont {Jarillo-Herrero}}]{cao18b}%
  \BibitemOpen
  \bibfield  {author} {\bibinfo {author} {\bibfnamefont {Y.}~\bibnamefont
  {Cao}}, \bibinfo {author} {\bibfnamefont {V.}~\bibnamefont {Fatemi}},
  \bibinfo {author} {\bibfnamefont {S.}~\bibnamefont {Fang}}, \bibinfo {author}
  {\bibfnamefont {K.}~\bibnamefont {Watanabe}}, \bibinfo {author}
  {\bibfnamefont {T.}~\bibnamefont {Taniguchi}}, \bibinfo {author}
  {\bibfnamefont {E.}~\bibnamefont {Kaxiras}}, \ and\ \bibinfo {author}
  {\bibfnamefont {P.}~\bibnamefont {Jarillo-Herrero}},\ }\href@noop {}
  {\bibfield  {journal} {\bibinfo  {journal} {Nature}\ }\textbf {\bibinfo
  {volume} {556}},\ \bibinfo {pages} {43} (\bibinfo {year}
  {2018}{\natexlab{b}})}\BibitemShut {NoStop}%
\bibitem [{\citenamefont {Lu}\ \emph {et~al.}(2019{\natexlab{a}})\citenamefont
  {Lu}, \citenamefont {Stepanov}, \citenamefont {Yang}, \citenamefont {Xie},
  \citenamefont {Aamir}, \citenamefont {Das}, \citenamefont {Urgell},
  \citenamefont {Watanabe}, \citenamefont {Taniguchi}, \citenamefont {Zhang}
  \emph {et~al.}}]{lu2019superconductors}%
  \BibitemOpen
  \bibfield  {author} {\bibinfo {author} {\bibfnamefont {X.}~\bibnamefont
  {Lu}}, \bibinfo {author} {\bibfnamefont {P.}~\bibnamefont {Stepanov}},
  \bibinfo {author} {\bibfnamefont {W.}~\bibnamefont {Yang}}, \bibinfo {author}
  {\bibfnamefont {M.}~\bibnamefont {Xie}}, \bibinfo {author} {\bibfnamefont
  {M.~A.}\ \bibnamefont {Aamir}}, \bibinfo {author} {\bibfnamefont
  {I.}~\bibnamefont {Das}}, \bibinfo {author} {\bibfnamefont {C.}~\bibnamefont
  {Urgell}}, \bibinfo {author} {\bibfnamefont {K.}~\bibnamefont {Watanabe}},
  \bibinfo {author} {\bibfnamefont {T.}~\bibnamefont {Taniguchi}}, \bibinfo
  {author} {\bibfnamefont {G.}~\bibnamefont {Zhang}},  \emph {et~al.},\
  }\href@noop {} {\bibfield  {journal} {\bibinfo  {journal} {Nature}\ }\textbf
  {\bibinfo {volume} {574}},\ \bibinfo {pages} {653} (\bibinfo {year}
  {2019}{\natexlab{a}})}\BibitemShut {NoStop}%
\bibitem [{\citenamefont {Stepanov}\ \emph {et~al.}(2020)\citenamefont
  {Stepanov}, \citenamefont {Das}, \citenamefont {Lu}, \citenamefont
  {Fahimniya}, \citenamefont {Watanabe}, \citenamefont {Taniguchi},
  \citenamefont {Koppens}, \citenamefont {Lischner}, \citenamefont {Levitov},\
  and\ \citenamefont {Efetov}}]{stepanov2019interplay}%
  \BibitemOpen
  \bibfield  {author} {\bibinfo {author} {\bibfnamefont {P.}~\bibnamefont
  {Stepanov}}, \bibinfo {author} {\bibfnamefont {I.}~\bibnamefont {Das}},
  \bibinfo {author} {\bibfnamefont {X.}~\bibnamefont {Lu}}, \bibinfo {author}
  {\bibfnamefont {A.}~\bibnamefont {Fahimniya}}, \bibinfo {author}
  {\bibfnamefont {K.}~\bibnamefont {Watanabe}}, \bibinfo {author}
  {\bibfnamefont {T.}~\bibnamefont {Taniguchi}}, \bibinfo {author}
  {\bibfnamefont {F.~H.}\ \bibnamefont {Koppens}}, \bibinfo {author}
  {\bibfnamefont {J.}~\bibnamefont {Lischner}}, \bibinfo {author}
  {\bibfnamefont {L.}~\bibnamefont {Levitov}}, \ and\ \bibinfo {author}
  {\bibfnamefont {D.~K.}\ \bibnamefont {Efetov}},\ }\href@noop {} {\bibfield
  {journal} {\bibinfo  {journal} {Nature}\ }\textbf {\bibinfo {volume} {583}},\
  \bibinfo {pages} {375} (\bibinfo {year} {2020})}\BibitemShut {NoStop}%
\bibitem [{\citenamefont {Rozhkov}\ \emph {et~al.}(2016)\citenamefont
  {Rozhkov}, \citenamefont {Sboychakov}, \citenamefont {Rakhmanov},\ and\
  \citenamefont {Nori}}]{BiGraphrev}%
  \BibitemOpen
  \bibfield  {author} {\bibinfo {author} {\bibfnamefont {A.}~\bibnamefont
  {Rozhkov}}, \bibinfo {author} {\bibfnamefont {A.}~\bibnamefont {Sboychakov}},
  \bibinfo {author} {\bibfnamefont {A.}~\bibnamefont {Rakhmanov}}, \ and\
  \bibinfo {author} {\bibfnamefont {F.}~\bibnamefont {Nori}},\ }\href {\doibase
  https://doi.org/10.1016/j.physrep.2016.07.003} {\bibfield  {journal}
  {\bibinfo  {journal} {Physics Reports}\ }\textbf {\bibinfo {volume} {648}},\
  \bibinfo {pages} {1 } (\bibinfo {year} {2016})}\BibitemShut {NoStop}%
\bibitem [{\citenamefont {Shen}\ \emph {et~al.}(2020)\citenamefont {Shen},
  \citenamefont {Chu}, \citenamefont {Wu}, \citenamefont {Li}, \citenamefont
  {Wang}, \citenamefont {Zhao}, \citenamefont {Tang}, \citenamefont {Liu},
  \citenamefont {Tian}, \citenamefont {Watanabe} \emph
  {et~al.}}]{shen2019observation}%
  \BibitemOpen
  \bibfield  {author} {\bibinfo {author} {\bibfnamefont {C.}~\bibnamefont
  {Shen}}, \bibinfo {author} {\bibfnamefont {Y.}~\bibnamefont {Chu}}, \bibinfo
  {author} {\bibfnamefont {Q.}~\bibnamefont {Wu}}, \bibinfo {author}
  {\bibfnamefont {N.}~\bibnamefont {Li}}, \bibinfo {author} {\bibfnamefont
  {S.}~\bibnamefont {Wang}}, \bibinfo {author} {\bibfnamefont {Y.}~\bibnamefont
  {Zhao}}, \bibinfo {author} {\bibfnamefont {J.}~\bibnamefont {Tang}}, \bibinfo
  {author} {\bibfnamefont {J.}~\bibnamefont {Liu}}, \bibinfo {author}
  {\bibfnamefont {J.}~\bibnamefont {Tian}}, \bibinfo {author} {\bibfnamefont
  {K.}~\bibnamefont {Watanabe}},  \emph {et~al.},\ }\href@noop {} {\bibfield
  {journal} {\bibinfo  {journal} {Nature Phys.}\ }\textbf {\bibinfo {volume}
  {16}},\ \bibinfo {pages} {520} (\bibinfo {year} {2020})}\BibitemShut
  {NoStop}%
\bibitem [{\citenamefont {Kerelsky}\ \emph {et~al.}(2021)\citenamefont
  {Kerelsky}, \citenamefont {Rubio-Verd{\'u}}, \citenamefont {Xian},
  \citenamefont {Kennes}, \citenamefont {Halbertal}, \citenamefont {Finney},
  \citenamefont {Song}, \citenamefont {Turkel}, \citenamefont {Wang},
  \citenamefont {Watanabe}, \citenamefont {Taniguchi}, \citenamefont {Hone},
  \citenamefont {Dean}, \citenamefont {Basov}, \citenamefont {Rubio},\ and\
  \citenamefont {Pasupathy}}]{Kerelskye2017366118}%
  \BibitemOpen
  \bibfield  {author} {\bibinfo {author} {\bibfnamefont {A.}~\bibnamefont
  {Kerelsky}}, \bibinfo {author} {\bibfnamefont {C.}~\bibnamefont
  {Rubio-Verd{\'u}}}, \bibinfo {author} {\bibfnamefont {L.}~\bibnamefont
  {Xian}}, \bibinfo {author} {\bibfnamefont {D.~M.}\ \bibnamefont {Kennes}},
  \bibinfo {author} {\bibfnamefont {D.}~\bibnamefont {Halbertal}}, \bibinfo
  {author} {\bibfnamefont {N.}~\bibnamefont {Finney}}, \bibinfo {author}
  {\bibfnamefont {L.}~\bibnamefont {Song}}, \bibinfo {author} {\bibfnamefont
  {S.}~\bibnamefont {Turkel}}, \bibinfo {author} {\bibfnamefont
  {L.}~\bibnamefont {Wang}}, \bibinfo {author} {\bibfnamefont {K.}~\bibnamefont
  {Watanabe}}, \bibinfo {author} {\bibfnamefont {T.}~\bibnamefont {Taniguchi}},
  \bibinfo {author} {\bibfnamefont {J.}~\bibnamefont {Hone}}, \bibinfo {author}
  {\bibfnamefont {C.}~\bibnamefont {Dean}}, \bibinfo {author} {\bibfnamefont
  {D.~N.}\ \bibnamefont {Basov}}, \bibinfo {author} {\bibfnamefont
  {A.}~\bibnamefont {Rubio}}, \ and\ \bibinfo {author} {\bibfnamefont {A.~N.}\
  \bibnamefont {Pasupathy}},\ }\href {\doibase 10.1073/pnas.2017366118}
  {\bibfield  {journal} {\bibinfo  {journal} {Proceedings of the National
  Academy of Sciences}\ }\textbf {\bibinfo {volume} {118}} (\bibinfo {year}
  {2021}),\ 10.1073/pnas.2017366118}\BibitemShut {NoStop}%
\bibitem [{\citenamefont {Rubio-Verdú}\ \emph {et~al.}(2020)\citenamefont
  {Rubio-Verdú}, \citenamefont {Turkel}, \citenamefont {Song}, \citenamefont
  {Klebl}, \citenamefont {Samajdar}, \citenamefont {Scheurer}, \citenamefont
  {Venderbos}, \citenamefont {Watanabe}, \citenamefont {Taniguchi},
  \citenamefont {Ochoa}, \citenamefont {Xian}, \citenamefont {Kennes},
  \citenamefont {Fernandes}, \citenamefont {Ángel Rubio},\ and\ \citenamefont
  {Pasupathy}}]{rubioverdu2020universal}%
  \BibitemOpen
  \bibfield  {author} {\bibinfo {author} {\bibfnamefont {C.}~\bibnamefont
  {Rubio-Verdú}}, \bibinfo {author} {\bibfnamefont {S.}~\bibnamefont
  {Turkel}}, \bibinfo {author} {\bibfnamefont {L.}~\bibnamefont {Song}},
  \bibinfo {author} {\bibfnamefont {L.}~\bibnamefont {Klebl}}, \bibinfo
  {author} {\bibfnamefont {R.}~\bibnamefont {Samajdar}}, \bibinfo {author}
  {\bibfnamefont {M.~S.}\ \bibnamefont {Scheurer}}, \bibinfo {author}
  {\bibfnamefont {J.~W.~F.}\ \bibnamefont {Venderbos}}, \bibinfo {author}
  {\bibfnamefont {K.}~\bibnamefont {Watanabe}}, \bibinfo {author}
  {\bibfnamefont {T.}~\bibnamefont {Taniguchi}}, \bibinfo {author}
  {\bibfnamefont {H.}~\bibnamefont {Ochoa}}, \bibinfo {author} {\bibfnamefont
  {L.}~\bibnamefont {Xian}}, \bibinfo {author} {\bibfnamefont {D.}~\bibnamefont
  {Kennes}}, \bibinfo {author} {\bibfnamefont {R.~M.}\ \bibnamefont
  {Fernandes}}, \bibinfo {author} {\bibnamefont {Ángel Rubio}}, \ and\
  \bibinfo {author} {\bibfnamefont {A.~N.}\ \bibnamefont {Pasupathy}},\
  }\href@noop {} {\  (\bibinfo {year} {2020})},\ \Eprint
  {http://arxiv.org/abs/2009.11645} {arXiv:2009.11645 [cond-mat.str-el]}
  \BibitemShut {NoStop}%
\bibitem [{\citenamefont {Arora}\ \emph {et~al.}(2020)\citenamefont {Arora},
  \citenamefont {Polski}, \citenamefont {Zhang}, \citenamefont {Thomson},
  \citenamefont {Choi}, \citenamefont {Kim}, \citenamefont {Lin}, \citenamefont
  {Wilson}, \citenamefont {Xu}, \citenamefont {Chu} \emph
  {et~al.}}]{arora2020superconductivity}%
  \BibitemOpen
  \bibfield  {author} {\bibinfo {author} {\bibfnamefont {H.~S.}\ \bibnamefont
  {Arora}}, \bibinfo {author} {\bibfnamefont {R.}~\bibnamefont {Polski}},
  \bibinfo {author} {\bibfnamefont {Y.}~\bibnamefont {Zhang}}, \bibinfo
  {author} {\bibfnamefont {A.}~\bibnamefont {Thomson}}, \bibinfo {author}
  {\bibfnamefont {Y.}~\bibnamefont {Choi}}, \bibinfo {author} {\bibfnamefont
  {H.}~\bibnamefont {Kim}}, \bibinfo {author} {\bibfnamefont {Z.}~\bibnamefont
  {Lin}}, \bibinfo {author} {\bibfnamefont {I.~Z.}\ \bibnamefont {Wilson}},
  \bibinfo {author} {\bibfnamefont {X.}~\bibnamefont {Xu}}, \bibinfo {author}
  {\bibfnamefont {J.-H.}\ \bibnamefont {Chu}},  \emph {et~al.},\ }\href@noop {}
  {\bibfield  {journal} {\bibinfo  {journal} {Nature}\ }\textbf {\bibinfo
  {volume} {583}},\ \bibinfo {pages} {379} (\bibinfo {year}
  {2020})}\BibitemShut {NoStop}%
\bibitem [{\citenamefont {Chen}\ \emph
  {et~al.}(2020{\natexlab{a}})\citenamefont {Chen}, \citenamefont {Sharpe},
  \citenamefont {Fox}, \citenamefont {Zhang}, \citenamefont {Wang},
  \citenamefont {Jiang}, \citenamefont {Lyu}, \citenamefont {Li}, \citenamefont
  {Watanabe}, \citenamefont {Taniguchi} \emph {et~al.}}]{chen2020nature}%
  \BibitemOpen
  \bibfield  {author} {\bibinfo {author} {\bibfnamefont {G.}~\bibnamefont
  {Chen}}, \bibinfo {author} {\bibfnamefont {A.~L.}\ \bibnamefont {Sharpe}},
  \bibinfo {author} {\bibfnamefont {E.~J.}\ \bibnamefont {Fox}}, \bibinfo
  {author} {\bibfnamefont {Y.-H.}\ \bibnamefont {Zhang}}, \bibinfo {author}
  {\bibfnamefont {S.}~\bibnamefont {Wang}}, \bibinfo {author} {\bibfnamefont
  {L.}~\bibnamefont {Jiang}}, \bibinfo {author} {\bibfnamefont
  {B.}~\bibnamefont {Lyu}}, \bibinfo {author} {\bibfnamefont {H.}~\bibnamefont
  {Li}}, \bibinfo {author} {\bibfnamefont {K.}~\bibnamefont {Watanabe}},
  \bibinfo {author} {\bibfnamefont {T.}~\bibnamefont {Taniguchi}},  \emph
  {et~al.},\ }\href@noop {} {\bibfield  {journal} {\bibinfo  {journal}
  {Nature}\ }\textbf {\bibinfo {volume} {579}},\ \bibinfo {pages} {56}
  (\bibinfo {year} {2020}{\natexlab{a}})}\BibitemShut {NoStop}%
\bibitem [{\citenamefont {Liu}\ \emph {et~al.}(2020{\natexlab{a}})\citenamefont
  {Liu}, \citenamefont {Hao}, \citenamefont {Khalaf}, \citenamefont {Lee},
  \citenamefont {Ronen}, \citenamefont {Yoo}, \citenamefont {Najafabadi},
  \citenamefont {Watanabe}, \citenamefont {Taniguchi}, \citenamefont
  {Vishwanath} \emph {et~al.}}]{liu2019spin}%
  \BibitemOpen
  \bibfield  {author} {\bibinfo {author} {\bibfnamefont {X.}~\bibnamefont
  {Liu}}, \bibinfo {author} {\bibfnamefont {Z.}~\bibnamefont {Hao}}, \bibinfo
  {author} {\bibfnamefont {E.}~\bibnamefont {Khalaf}}, \bibinfo {author}
  {\bibfnamefont {J.~Y.}\ \bibnamefont {Lee}}, \bibinfo {author} {\bibfnamefont
  {Y.}~\bibnamefont {Ronen}}, \bibinfo {author} {\bibfnamefont
  {H.}~\bibnamefont {Yoo}}, \bibinfo {author} {\bibfnamefont {D.~H.}\
  \bibnamefont {Najafabadi}}, \bibinfo {author} {\bibfnamefont
  {K.}~\bibnamefont {Watanabe}}, \bibinfo {author} {\bibfnamefont
  {T.}~\bibnamefont {Taniguchi}}, \bibinfo {author} {\bibfnamefont
  {A.}~\bibnamefont {Vishwanath}},  \emph {et~al.},\ }\href@noop {} {\bibfield
  {journal} {\bibinfo  {journal} {Nature}\ }\textbf {\bibinfo {volume} {583}},\
  \bibinfo {pages} {221} (\bibinfo {year} {2020}{\natexlab{a}})}\BibitemShut
  {NoStop}%
\bibitem [{\citenamefont {Cao}\ \emph {et~al.}(2020{\natexlab{a}})\citenamefont
  {Cao}, \citenamefont {Rodan-Legrain}, \citenamefont {Rubies-Bigorda},
  \citenamefont {Park}, \citenamefont {Watanabe}, \citenamefont {Taniguchi},\
  and\ \citenamefont {Jarillo-Herrero}}]{cao2019electric}%
  \BibitemOpen
  \bibfield  {author} {\bibinfo {author} {\bibfnamefont {Y.}~\bibnamefont
  {Cao}}, \bibinfo {author} {\bibfnamefont {D.}~\bibnamefont {Rodan-Legrain}},
  \bibinfo {author} {\bibfnamefont {O.}~\bibnamefont {Rubies-Bigorda}},
  \bibinfo {author} {\bibfnamefont {J.~M.}\ \bibnamefont {Park}}, \bibinfo
  {author} {\bibfnamefont {K.}~\bibnamefont {Watanabe}}, \bibinfo {author}
  {\bibfnamefont {T.}~\bibnamefont {Taniguchi}}, \ and\ \bibinfo {author}
  {\bibfnamefont {P.}~\bibnamefont {Jarillo-Herrero}},\ }\href@noop {}
  {\bibfield  {journal} {\bibinfo  {journal} {Nature}\ }\textbf {\bibinfo
  {volume} {583}},\ \bibinfo {pages} {215} (\bibinfo {year}
  {2020}{\natexlab{a}})}\BibitemShut {NoStop}%
\bibitem [{\citenamefont {He}\ \emph {et~al.}(2020)\citenamefont {He},
  \citenamefont {Li}, \citenamefont {Cai}, \citenamefont {Liu}, \citenamefont
  {Watanabe}, \citenamefont {Taniguchi}, \citenamefont {Xu},\ and\
  \citenamefont {Yankowitz}}]{he2020}%
  \BibitemOpen
  \bibfield  {author} {\bibinfo {author} {\bibfnamefont {M.}~\bibnamefont
  {He}}, \bibinfo {author} {\bibfnamefont {Y.}~\bibnamefont {Li}}, \bibinfo
  {author} {\bibfnamefont {J.}~\bibnamefont {Cai}}, \bibinfo {author}
  {\bibfnamefont {Y.}~\bibnamefont {Liu}}, \bibinfo {author} {\bibfnamefont
  {K.}~\bibnamefont {Watanabe}}, \bibinfo {author} {\bibfnamefont
  {T.}~\bibnamefont {Taniguchi}}, \bibinfo {author} {\bibfnamefont
  {X.}~\bibnamefont {Xu}}, \ and\ \bibinfo {author} {\bibfnamefont
  {M.}~\bibnamefont {Yankowitz}},\ }\href@noop {} {\bibfield  {journal}
  {\bibinfo  {journal} {Nat. Phys.}\ } (\bibinfo {year} {2020})}\BibitemShut
  {NoStop}%
\bibitem [{\citenamefont {Wang}\ \emph {et~al.}(2020)\citenamefont {Wang},
  \citenamefont {Shih}, \citenamefont {Ghiotto}, \citenamefont {Xian},
  \citenamefont {Rhodes}, \citenamefont {Tan}, \citenamefont {Claassen},
  \citenamefont {Kennes}, \citenamefont {Bai}, \citenamefont {Kim} \emph
  {et~al.}}]{Wang19TMD}%
  \BibitemOpen
  \bibfield  {author} {\bibinfo {author} {\bibfnamefont {L.}~\bibnamefont
  {Wang}}, \bibinfo {author} {\bibfnamefont {E.-M.}\ \bibnamefont {Shih}},
  \bibinfo {author} {\bibfnamefont {A.}~\bibnamefont {Ghiotto}}, \bibinfo
  {author} {\bibfnamefont {L.}~\bibnamefont {Xian}}, \bibinfo {author}
  {\bibfnamefont {D.~A.}\ \bibnamefont {Rhodes}}, \bibinfo {author}
  {\bibfnamefont {C.}~\bibnamefont {Tan}}, \bibinfo {author} {\bibfnamefont
  {M.}~\bibnamefont {Claassen}}, \bibinfo {author} {\bibfnamefont {D.~M.}\
  \bibnamefont {Kennes}}, \bibinfo {author} {\bibfnamefont {Y.}~\bibnamefont
  {Bai}}, \bibinfo {author} {\bibfnamefont {B.}~\bibnamefont {Kim}},  \emph
  {et~al.},\ }\href@noop {} {\bibfield  {journal} {\bibinfo  {journal} {Nat.
  Mater.}\ }\textbf {\bibinfo {volume} {19}},\ \bibinfo {pages} {861} (\bibinfo
  {year} {2020})}\BibitemShut {NoStop}%
\bibitem [{\citenamefont {Chen}\ \emph
  {et~al.}(2019{\natexlab{a}})\citenamefont {Chen}, \citenamefont {Jiang},
  \citenamefont {Wu}, \citenamefont {Lv}, \citenamefont {Li}, \citenamefont
  {Watanabe}, \citenamefont {Taniguchi}, \citenamefont {Shi}, \citenamefont
  {Zhang},\ and\ \citenamefont {Wang}}]{chen2018gate}%
  \BibitemOpen
  \bibfield  {author} {\bibinfo {author} {\bibfnamefont {G.}~\bibnamefont
  {Chen}}, \bibinfo {author} {\bibfnamefont {L.}~\bibnamefont {Jiang}},
  \bibinfo {author} {\bibfnamefont {S.}~\bibnamefont {Wu}}, \bibinfo {author}
  {\bibfnamefont {B.}~\bibnamefont {Lv}}, \bibinfo {author} {\bibfnamefont
  {H.}~\bibnamefont {Li}}, \bibinfo {author} {\bibfnamefont {K.}~\bibnamefont
  {Watanabe}}, \bibinfo {author} {\bibfnamefont {T.}~\bibnamefont {Taniguchi}},
  \bibinfo {author} {\bibfnamefont {Z.}~\bibnamefont {Shi}}, \bibinfo {author}
  {\bibfnamefont {Y.}~\bibnamefont {Zhang}}, \ and\ \bibinfo {author}
  {\bibfnamefont {F.}~\bibnamefont {Wang}},\ }\href@noop {} {\bibfield
  {journal} {\bibinfo  {journal} {Nature Phys.}\ }\textbf {\bibinfo {volume}
  {15}},\ \bibinfo {pages} {237} (\bibinfo {year}
  {2019}{\natexlab{a}})}\BibitemShut {NoStop}%
\bibitem [{\citenamefont {Chen}\ \emph
  {et~al.}(2019{\natexlab{b}})\citenamefont {Chen}, \citenamefont {Sharpe},
  \citenamefont {Gallagher}, \citenamefont {Rosen}, \citenamefont {Fox},
  \citenamefont {Jiang}, \citenamefont {Lyu}, \citenamefont {Li}, \citenamefont
  {Watanabe}, \citenamefont {Taniguchi} \emph {et~al.}}]{chen2019signatures}%
  \BibitemOpen
  \bibfield  {author} {\bibinfo {author} {\bibfnamefont {G.}~\bibnamefont
  {Chen}}, \bibinfo {author} {\bibfnamefont {A.~L.}\ \bibnamefont {Sharpe}},
  \bibinfo {author} {\bibfnamefont {P.}~\bibnamefont {Gallagher}}, \bibinfo
  {author} {\bibfnamefont {I.~T.}\ \bibnamefont {Rosen}}, \bibinfo {author}
  {\bibfnamefont {E.}~\bibnamefont {Fox}}, \bibinfo {author} {\bibfnamefont
  {L.}~\bibnamefont {Jiang}}, \bibinfo {author} {\bibfnamefont
  {B.}~\bibnamefont {Lyu}}, \bibinfo {author} {\bibfnamefont {H.}~\bibnamefont
  {Li}}, \bibinfo {author} {\bibfnamefont {K.}~\bibnamefont {Watanabe}},
  \bibinfo {author} {\bibfnamefont {T.}~\bibnamefont {Taniguchi}},  \emph
  {et~al.},\ }\href@noop {} {\bibfield  {journal} {\bibinfo  {journal}
  {Nature}\ }\textbf {\bibinfo {volume} {572}},\ \bibinfo {pages} {215}
  (\bibinfo {year} {2019}{\natexlab{b}})}\BibitemShut {NoStop}%
\bibitem [{\citenamefont {Lu}\ \emph {et~al.}(2019{\natexlab{b}})\citenamefont
  {Lu}, \citenamefont {Stepanov}, \citenamefont {Yang}, \citenamefont {Xie},
  \citenamefont {Aamir}, \citenamefont {Das}, \citenamefont {Urgell},
  \citenamefont {Watanabe}, \citenamefont {Taniguchi}, \citenamefont {Zhang},
  \citenamefont {Bachtold}, \citenamefont {MacDonald},\ and\ \citenamefont
  {Efetov}}]{lu19}%
  \BibitemOpen
  \bibfield  {author} {\bibinfo {author} {\bibfnamefont {X.}~\bibnamefont
  {Lu}}, \bibinfo {author} {\bibfnamefont {P.}~\bibnamefont {Stepanov}},
  \bibinfo {author} {\bibfnamefont {W.}~\bibnamefont {Yang}}, \bibinfo {author}
  {\bibfnamefont {M.}~\bibnamefont {Xie}}, \bibinfo {author} {\bibfnamefont
  {M.~A.}\ \bibnamefont {Aamir}}, \bibinfo {author} {\bibfnamefont
  {I.}~\bibnamefont {Das}}, \bibinfo {author} {\bibfnamefont {C.}~\bibnamefont
  {Urgell}}, \bibinfo {author} {\bibfnamefont {K.}~\bibnamefont {Watanabe}},
  \bibinfo {author} {\bibfnamefont {T.}~\bibnamefont {Taniguchi}}, \bibinfo
  {author} {\bibfnamefont {G.}~\bibnamefont {Zhang}}, \bibinfo {author}
  {\bibfnamefont {A.}~\bibnamefont {Bachtold}}, \bibinfo {author}
  {\bibfnamefont {A.~H.}\ \bibnamefont {MacDonald}}, \ and\ \bibinfo {author}
  {\bibfnamefont {D.~K.}\ \bibnamefont {Efetov}},\ }\href@noop {} {\bibfield
  {journal} {\bibinfo  {journal} {Nature}\ }\textbf {\bibinfo {volume} {574}},\
  \bibinfo {pages} {653} (\bibinfo {year} {2019}{\natexlab{b}})}\BibitemShut
  {NoStop}%
\bibitem [{\citenamefont {Cao}\ \emph {et~al.}(2020{\natexlab{b}})\citenamefont
  {Cao}, \citenamefont {Chowdhury}, \citenamefont {Rodan-Legrain},
  \citenamefont {Rubies-Bigorda}, \citenamefont {Watanabe}, \citenamefont
  {Taniguchi}, \citenamefont {Senthil},\ and\ \citenamefont
  {Jarillo-Herrero}}]{cao20a}%
  \BibitemOpen
  \bibfield  {author} {\bibinfo {author} {\bibfnamefont {Y.}~\bibnamefont
  {Cao}}, \bibinfo {author} {\bibfnamefont {D.}~\bibnamefont {Chowdhury}},
  \bibinfo {author} {\bibfnamefont {D.}~\bibnamefont {Rodan-Legrain}}, \bibinfo
  {author} {\bibfnamefont {O.}~\bibnamefont {Rubies-Bigorda}}, \bibinfo
  {author} {\bibfnamefont {K.}~\bibnamefont {Watanabe}}, \bibinfo {author}
  {\bibfnamefont {T.}~\bibnamefont {Taniguchi}}, \bibinfo {author}
  {\bibfnamefont {T.}~\bibnamefont {Senthil}}, \ and\ \bibinfo {author}
  {\bibfnamefont {P.}~\bibnamefont {Jarillo-Herrero}},\ }\href@noop {}
  {\bibfield  {journal} {\bibinfo  {journal} {Phys. Rev. Lett.}\ }\textbf
  {\bibinfo {volume} {124}},\ \bibinfo {pages} {076801} (\bibinfo {year}
  {2020}{\natexlab{b}})}\BibitemShut {NoStop}%
\bibitem [{\citenamefont {Sharpe}\ \emph {et~al.}(2019)\citenamefont {Sharpe},
  \citenamefont {Fox}, \citenamefont {Barnard}, \citenamefont {Finney},
  \citenamefont {Watanabe}, \citenamefont {Taniguchi}, \citenamefont
  {Kastner},\ and\ \citenamefont {Goldhaber-Gordon}}]{sharpe19}%
  \BibitemOpen
  \bibfield  {author} {\bibinfo {author} {\bibfnamefont {A.~L.}\ \bibnamefont
  {Sharpe}}, \bibinfo {author} {\bibfnamefont {E.~J.}\ \bibnamefont {Fox}},
  \bibinfo {author} {\bibfnamefont {A.~W.}\ \bibnamefont {Barnard}}, \bibinfo
  {author} {\bibfnamefont {J.}~\bibnamefont {Finney}}, \bibinfo {author}
  {\bibfnamefont {K.}~\bibnamefont {Watanabe}}, \bibinfo {author}
  {\bibfnamefont {T.}~\bibnamefont {Taniguchi}}, \bibinfo {author}
  {\bibfnamefont {M.~A.}\ \bibnamefont {Kastner}}, \ and\ \bibinfo {author}
  {\bibfnamefont {D.}~\bibnamefont {Goldhaber-Gordon}},\ }\href@noop {}
  {\bibfield  {journal} {\bibinfo  {journal} {Science}\ }\textbf {\bibinfo
  {volume} {365}},\ \bibinfo {pages} {605} (\bibinfo {year}
  {2019})}\BibitemShut {NoStop}%
\bibitem [{\citenamefont {Serlin}\ \emph {et~al.}(2020)\citenamefont {Serlin},
  \citenamefont {Tschirhart}, \citenamefont {Polshyn}, \citenamefont {Zhang},
  \citenamefont {Zhu}, \citenamefont {Watanabe}, \citenamefont {Taniguchi},
  \citenamefont {Balents},\ and\ \citenamefont {Young}}]{serlin20}%
  \BibitemOpen
  \bibfield  {author} {\bibinfo {author} {\bibfnamefont {M.}~\bibnamefont
  {Serlin}}, \bibinfo {author} {\bibfnamefont {C.~L.}\ \bibnamefont
  {Tschirhart}}, \bibinfo {author} {\bibfnamefont {H.}~\bibnamefont {Polshyn}},
  \bibinfo {author} {\bibfnamefont {Y.}~\bibnamefont {Zhang}}, \bibinfo
  {author} {\bibfnamefont {J.}~\bibnamefont {Zhu}}, \bibinfo {author}
  {\bibfnamefont {K.}~\bibnamefont {Watanabe}}, \bibinfo {author}
  {\bibfnamefont {T.}~\bibnamefont {Taniguchi}}, \bibinfo {author}
  {\bibfnamefont {L.}~\bibnamefont {Balents}}, \ and\ \bibinfo {author}
  {\bibfnamefont {A.~F.}\ \bibnamefont {Young}},\ }\href@noop {} {\bibfield
  {journal} {\bibinfo  {journal} {Science}\ }\textbf {\bibinfo {volume}
  {367}},\ \bibinfo {pages} {900} (\bibinfo {year} {2020})}\BibitemShut
  {NoStop}%
\bibitem [{\citenamefont {Wu}\ \emph {et~al.}(2021)\citenamefont {Wu},
  \citenamefont {Zhang}, \citenamefont {Watanabe}, \citenamefont {Taniguchi},\
  and\ \citenamefont {Andrei}}]{wu21}%
  \BibitemOpen
  \bibfield  {author} {\bibinfo {author} {\bibfnamefont {S.}~\bibnamefont
  {Wu}}, \bibinfo {author} {\bibfnamefont {Z.}~\bibnamefont {Zhang}}, \bibinfo
  {author} {\bibfnamefont {K.}~\bibnamefont {Watanabe}}, \bibinfo {author}
  {\bibfnamefont {T.}~\bibnamefont {Taniguchi}}, \ and\ \bibinfo {author}
  {\bibfnamefont {E.~Y.}\ \bibnamefont {Andrei}},\ }\href@noop {} {\bibfield
  {journal} {\bibinfo  {journal} {Nature Mat.}\ }\textbf {\bibinfo {volume}
  {20}},\ \bibinfo {pages} {488} (\bibinfo {year} {2021})}\BibitemShut
  {NoStop}%
\bibitem [{\citenamefont {Chen}\ \emph
  {et~al.}(2020{\natexlab{b}})\citenamefont {Chen}, \citenamefont {Sharpe},
  \citenamefont {Fox}, \citenamefont {Zhang}, \citenamefont {Wang},
  \citenamefont {Jiang}, \citenamefont {Lyu}, \citenamefont {Li}, \citenamefont
  {Watanabe}, \citenamefont {Taniguchi}, \citenamefont {Shi}, \citenamefont
  {Senthil}, \citenamefont {Goldhaber-Gordon}, \citenamefont {Zhang},\ and\
  \citenamefont {Wang}}]{chen20}%
  \BibitemOpen
  \bibfield  {author} {\bibinfo {author} {\bibfnamefont {G.}~\bibnamefont
  {Chen}}, \bibinfo {author} {\bibfnamefont {A.~L.}\ \bibnamefont {Sharpe}},
  \bibinfo {author} {\bibfnamefont {E.~J.}\ \bibnamefont {Fox}}, \bibinfo
  {author} {\bibfnamefont {Y.~H.}\ \bibnamefont {Zhang}}, \bibinfo {author}
  {\bibfnamefont {S.}~\bibnamefont {Wang}}, \bibinfo {author} {\bibfnamefont
  {L.}~\bibnamefont {Jiang}}, \bibinfo {author} {\bibfnamefont
  {B.}~\bibnamefont {Lyu}}, \bibinfo {author} {\bibfnamefont {H.}~\bibnamefont
  {Li}}, \bibinfo {author} {\bibfnamefont {K.}~\bibnamefont {Watanabe}},
  \bibinfo {author} {\bibfnamefont {T.}~\bibnamefont {Taniguchi}}, \bibinfo
  {author} {\bibfnamefont {Z.}~\bibnamefont {Shi}}, \bibinfo {author}
  {\bibfnamefont {T.}~\bibnamefont {Senthil}}, \bibinfo {author} {\bibfnamefont
  {D.}~\bibnamefont {Goldhaber-Gordon}}, \bibinfo {author} {\bibfnamefont
  {Y.}~\bibnamefont {Zhang}}, \ and\ \bibinfo {author} {\bibfnamefont
  {F.}~\bibnamefont {Wang}},\ }\href@noop {} {\bibfield  {journal} {\bibinfo
  {journal} {Nature}\ }\textbf {\bibinfo {volume} {579}},\ \bibinfo {pages}
  {56} (\bibinfo {year} {2020}{\natexlab{b}})}\BibitemShut {NoStop}%
\bibitem [{\citenamefont {Liu}\ \emph {et~al.}(2020{\natexlab{b}})\citenamefont
  {Liu}, \citenamefont {Abouelkomsan},\ and\ \citenamefont
  {Bergholtz}}]{liu20}%
  \BibitemOpen
  \bibfield  {author} {\bibinfo {author} {\bibfnamefont {Z.}~\bibnamefont
  {Liu}}, \bibinfo {author} {\bibfnamefont {A.}~\bibnamefont {Abouelkomsan}}, \
  and\ \bibinfo {author} {\bibfnamefont {E.~J.}\ \bibnamefont {Bergholtz}},\
  }\href@noop {} {\bibfield  {journal} {\bibinfo  {journal} {arXiv:2004.09522}\
  } (\bibinfo {year} {2020}{\natexlab{b}})}\BibitemShut {NoStop}%
\bibitem [{\citenamefont {Abouelkomsan}\ \emph {et~al.}(2019)\citenamefont
  {Abouelkomsan}, \citenamefont {Liu},\ and\ \citenamefont
  {Bergholtz}}]{abouelkomsan19}%
  \BibitemOpen
  \bibfield  {author} {\bibinfo {author} {\bibfnamefont {A.}~\bibnamefont
  {Abouelkomsan}}, \bibinfo {author} {\bibfnamefont {Z.}~\bibnamefont {Liu}}, \
  and\ \bibinfo {author} {\bibfnamefont {E.~J.}\ \bibnamefont {Bergholtz}},\
  }\href@noop {} {\bibfield  {journal} {\bibinfo  {journal} {Phys. Rev. Lett.}\
  }\textbf {\bibinfo {volume} {124}},\ \bibinfo {pages} {106803} (\bibinfo
  {year} {2019})}\BibitemShut {NoStop}%
\bibitem [{\citenamefont {Repellin}\ and\ \citenamefont
  {Senthil}(2019)}]{repellin19}%
  \BibitemOpen
  \bibfield  {author} {\bibinfo {author} {\bibfnamefont {C.}~\bibnamefont
  {Repellin}}\ and\ \bibinfo {author} {\bibfnamefont {T.}~\bibnamefont
  {Senthil}},\ }\href@noop {} {\bibfield  {journal} {\bibinfo  {journal}
  {arXiv:1912.11469}\ } (\bibinfo {year} {2019})}\BibitemShut {NoStop}%
\bibitem [{\citenamefont {Ledwith}\ \emph {et~al.}(2019)\citenamefont
  {Ledwith}, \citenamefont {Tarnopolsky}, \citenamefont {Khalaf},\ and\
  \citenamefont {Vishwanath}}]{ledwith19}%
  \BibitemOpen
  \bibfield  {author} {\bibinfo {author} {\bibfnamefont {P.~J.}\ \bibnamefont
  {Ledwith}}, \bibinfo {author} {\bibfnamefont {G.}~\bibnamefont
  {Tarnopolsky}}, \bibinfo {author} {\bibfnamefont {E.}~\bibnamefont {Khalaf}},
  \ and\ \bibinfo {author} {\bibfnamefont {A.}~\bibnamefont {Vishwanath}},\
  }\href@noop {} {\bibfield  {journal} {\bibinfo  {journal} {arXiv:1912.09634}\
  } (\bibinfo {year} {2019})}\BibitemShut {NoStop}%
\bibitem [{\citenamefont {Wu}\ \emph {et~al.}(2018)\citenamefont {Wu},
  \citenamefont {Lovorn}, \citenamefont {Tutuc}, \citenamefont {Martin},\ and\
  \citenamefont {MacDonald}}]{wu18}%
  \BibitemOpen
  \bibfield  {author} {\bibinfo {author} {\bibfnamefont {F.}~\bibnamefont
  {Wu}}, \bibinfo {author} {\bibfnamefont {T.}~\bibnamefont {Lovorn}}, \bibinfo
  {author} {\bibfnamefont {E.}~\bibnamefont {Tutuc}}, \bibinfo {author}
  {\bibfnamefont {I.}~\bibnamefont {Martin}}, \ and\ \bibinfo {author}
  {\bibfnamefont {A.~H.}\ \bibnamefont {MacDonald}},\ }\href@noop {} {\bibfield
   {journal} {\bibinfo  {journal} {Phys. Rev. Lett.}\ }\textbf {\bibinfo
  {volume} {122}},\ \bibinfo {pages} {086402} (\bibinfo {year} {2018})},\
  \Eprint {http://arxiv.org/abs/arXiv:1807.03311} {arXiv:1807.03311}
  \BibitemShut {NoStop}%
\bibitem [{\citenamefont {Li}\ \emph {et~al.}(2021)\citenamefont {Li},
  \citenamefont {Jiang}, \citenamefont {Shen}, \citenamefont {Zhang},
  \citenamefont {Li}, \citenamefont {Devakul}, \citenamefont {Watanabe},
  \citenamefont {Taniguchi}, \citenamefont {Fu}, \citenamefont {Shan},\ and\
  \citenamefont {Mak}}]{li21}%
  \BibitemOpen
  \bibfield  {author} {\bibinfo {author} {\bibfnamefont {T.}~\bibnamefont
  {Li}}, \bibinfo {author} {\bibfnamefont {S.}~\bibnamefont {Jiang}}, \bibinfo
  {author} {\bibfnamefont {B.}~\bibnamefont {Shen}}, \bibinfo {author}
  {\bibfnamefont {Y.}~\bibnamefont {Zhang}}, \bibinfo {author} {\bibfnamefont
  {L.}~\bibnamefont {Li}}, \bibinfo {author} {\bibfnamefont {T.}~\bibnamefont
  {Devakul}}, \bibinfo {author} {\bibfnamefont {K.}~\bibnamefont {Watanabe}},
  \bibinfo {author} {\bibfnamefont {T.}~\bibnamefont {Taniguchi}}, \bibinfo
  {author} {\bibfnamefont {L.}~\bibnamefont {Fu}}, \bibinfo {author}
  {\bibfnamefont {J.}~\bibnamefont {Shan}}, \ and\ \bibinfo {author}
  {\bibfnamefont {K.~F.}\ \bibnamefont {Mak}},\ }\href@noop {} {\bibfield
  {journal} {\bibinfo  {journal} {arXiv:2107.01796}\ } (\bibinfo {year}
  {2021})}\BibitemShut {NoStop}%
\bibitem [{\citenamefont {Zhang}\ \emph {et~al.}(2021)\citenamefont {Zhang},
  \citenamefont {Devakul},\ and\ \citenamefont {Fu}}]{zhang21}%
  \BibitemOpen
  \bibfield  {author} {\bibinfo {author} {\bibfnamefont {Y.}~\bibnamefont
  {Zhang}}, \bibinfo {author} {\bibfnamefont {T.}~\bibnamefont {Devakul}}, \
  and\ \bibinfo {author} {\bibfnamefont {L.}~\bibnamefont {Fu}},\ }\href@noop
  {} {\bibfield  {journal} {\bibinfo  {journal} {arXiv:2107.02167}\ } (\bibinfo
  {year} {2021})}\BibitemShut {NoStop}%
\bibitem [{\citenamefont {Xie}\ \emph {et~al.}(2021)\citenamefont {Xie},
  \citenamefont {Zhang}, \citenamefont {Hu}, \citenamefont {Mak},\ and\
  \citenamefont {Law}}]{xie21}%
  \BibitemOpen
  \bibfield  {author} {\bibinfo {author} {\bibfnamefont {Y.~M.}\ \bibnamefont
  {Xie}}, \bibinfo {author} {\bibfnamefont {C.~P.}\ \bibnamefont {Zhang}},
  \bibinfo {author} {\bibfnamefont {J.~X.}\ \bibnamefont {Hu}}, \bibinfo
  {author} {\bibfnamefont {K.~F.}\ \bibnamefont {Mak}}, \ and\ \bibinfo
  {author} {\bibfnamefont {K.~T.}\ \bibnamefont {Law}},\ }\href@noop {}
  {\bibfield  {journal} {\bibinfo  {journal} {arXiv:2106.13991}\ } (\bibinfo
  {year} {2021})}\BibitemShut {NoStop}%
\bibitem [{\citenamefont {Zhang}\ \emph {et~al.}(2015)\citenamefont {Zhang},
  \citenamefont {Zhu}, \citenamefont {Wang}, \citenamefont {Feng},
  \citenamefont {Qiao}, \citenamefont {Wen}, \citenamefont {Chen},
  \citenamefont {Cui}, \citenamefont {Zhang}, \citenamefont {Cai} \emph
  {et~al.}}]{zhang2015controlled}%
  \BibitemOpen
  \bibfield  {author} {\bibinfo {author} {\bibfnamefont {M.}~\bibnamefont
  {Zhang}}, \bibinfo {author} {\bibfnamefont {Y.}~\bibnamefont {Zhu}}, \bibinfo
  {author} {\bibfnamefont {X.}~\bibnamefont {Wang}}, \bibinfo {author}
  {\bibfnamefont {Q.}~\bibnamefont {Feng}}, \bibinfo {author} {\bibfnamefont
  {S.}~\bibnamefont {Qiao}}, \bibinfo {author} {\bibfnamefont {W.}~\bibnamefont
  {Wen}}, \bibinfo {author} {\bibfnamefont {Y.}~\bibnamefont {Chen}}, \bibinfo
  {author} {\bibfnamefont {M.}~\bibnamefont {Cui}}, \bibinfo {author}
  {\bibfnamefont {J.}~\bibnamefont {Zhang}}, \bibinfo {author} {\bibfnamefont
  {C.}~\bibnamefont {Cai}},  \emph {et~al.},\ }\href@noop {} {\bibfield
  {journal} {\bibinfo  {journal} {Journal of the American Chemical Society}\
  }\textbf {\bibinfo {volume} {137}},\ \bibinfo {pages} {7051} (\bibinfo {year}
  {2015})}\BibitemShut {NoStop}%
\bibitem [{\citenamefont {Xian}\ \emph {et~al.}(2020)\citenamefont {Xian},
  \citenamefont {Claassen}, \citenamefont {Kiese}, \citenamefont {Scherer},
  \citenamefont {Trebst}, \citenamefont {Kennes},\ and\ \citenamefont
  {Rubio}}]{xian20}%
  \BibitemOpen
  \bibfield  {author} {\bibinfo {author} {\bibfnamefont {L.}~\bibnamefont
  {Xian}}, \bibinfo {author} {\bibfnamefont {M.}~\bibnamefont {Claassen}},
  \bibinfo {author} {\bibfnamefont {D.}~\bibnamefont {Kiese}}, \bibinfo
  {author} {\bibfnamefont {M.~M.}\ \bibnamefont {Scherer}}, \bibinfo {author}
  {\bibfnamefont {S.}~\bibnamefont {Trebst}}, \bibinfo {author} {\bibfnamefont
  {D.~M.}\ \bibnamefont {Kennes}}, \ and\ \bibinfo {author} {\bibfnamefont
  {A.}~\bibnamefont {Rubio}},\ }\href@noop {} {\bibfield  {journal} {\bibinfo
  {journal} {arXiv:2004.02964}\ } (\bibinfo {year} {2020})}\BibitemShut
  {NoStop}%
\bibitem [{\citenamefont {Angeli}\ and\ \citenamefont
  {MacDonald}(2021)}]{angeli20}%
  \BibitemOpen
  \bibfield  {author} {\bibinfo {author} {\bibfnamefont {M.}~\bibnamefont
  {Angeli}}\ and\ \bibinfo {author} {\bibfnamefont {A.~H.}\ \bibnamefont
  {MacDonald}},\ }\href@noop {} {\bibfield  {journal} {\bibinfo  {journal}
  {Proc. Nat. Acad. Sci.}\ }\textbf {\bibinfo {volume} {118}},\ \bibinfo
  {pages} {e2021826118} (\bibinfo {year} {2021})}\BibitemShut {NoStop}%
\bibitem [{\citenamefont {Tang}\ \emph {et~al.}(2010)\citenamefont {Tang},
  \citenamefont {Mei},\ and\ \citenamefont {Wen}}]{tang10}%
  \BibitemOpen
  \bibfield  {author} {\bibinfo {author} {\bibfnamefont {E.}~\bibnamefont
  {Tang}}, \bibinfo {author} {\bibfnamefont {J.~W.}\ \bibnamefont {Mei}}, \
  and\ \bibinfo {author} {\bibfnamefont {X.~G.}\ \bibnamefont {Wen}},\
  }\href@noop {} {\bibfield  {journal} {\bibinfo  {journal} {Phys. Rev. Lett.}\
  }\textbf {\bibinfo {volume} {106}},\ \bibinfo {pages} {236802} (\bibinfo
  {year} {2010})}\BibitemShut {NoStop}%
\bibitem [{\citenamefont {Wu}\ \emph {et~al.}(2011)\citenamefont {Wu},
  \citenamefont {Bernevig},\ and\ \citenamefont {Regnault}}]{wu11}%
  \BibitemOpen
  \bibfield  {author} {\bibinfo {author} {\bibfnamefont {Y.~L.}\ \bibnamefont
  {Wu}}, \bibinfo {author} {\bibfnamefont {B.~A.}\ \bibnamefont {Bernevig}}, \
  and\ \bibinfo {author} {\bibfnamefont {N.}~\bibnamefont {Regnault}},\
  }\href@noop {} {\bibfield  {journal} {\bibinfo  {journal} {Phys. Rev. B}\
  }\textbf {\bibinfo {volume} {85}},\ \bibinfo {pages} {075116} (\bibinfo
  {year} {2011})}\BibitemShut {NoStop}%
\bibitem [{\citenamefont {Parameswaran}\ \emph {et~al.}(2011)\citenamefont
  {Parameswaran}, \citenamefont {Roy},\ and\ \citenamefont
  {Sondhi}}]{parameswaran11}%
  \BibitemOpen
  \bibfield  {author} {\bibinfo {author} {\bibfnamefont {S.~A.}\ \bibnamefont
  {Parameswaran}}, \bibinfo {author} {\bibfnamefont {R.}~\bibnamefont {Roy}}, \
  and\ \bibinfo {author} {\bibfnamefont {S.~L.}\ \bibnamefont {Sondhi}},\
  }\href@noop {} {\bibfield  {journal} {\bibinfo  {journal} {Phys. Rev. B}\
  }\textbf {\bibinfo {volume} {85}},\ \bibinfo {pages} {241308} (\bibinfo
  {year} {2011})}\BibitemShut {NoStop}%
\bibitem [{\citenamefont {Claassen}\ \emph {et~al.}(2015)\citenamefont
  {Claassen}, \citenamefont {Lee}, \citenamefont {Thomale}, \citenamefont
  {Qi},\ and\ \citenamefont {Devereaux}}]{claassen15}%
  \BibitemOpen
  \bibfield  {author} {\bibinfo {author} {\bibfnamefont {M.}~\bibnamefont
  {Claassen}}, \bibinfo {author} {\bibfnamefont {C.~H.}\ \bibnamefont {Lee}},
  \bibinfo {author} {\bibfnamefont {R.}~\bibnamefont {Thomale}}, \bibinfo
  {author} {\bibfnamefont {X.~L.}\ \bibnamefont {Qi}}, \ and\ \bibinfo {author}
  {\bibfnamefont {T.~P.}\ \bibnamefont {Devereaux}},\ }\href@noop {} {\bibfield
   {journal} {\bibinfo  {journal} {Phys. Rev. Lett.}\ }\textbf {\bibinfo
  {volume} {114}},\ \bibinfo {pages} {236802} (\bibinfo {year}
  {2015})}\BibitemShut {NoStop}%
\bibitem [{\citenamefont {Jackson}\ \emph {et~al.}(2014)\citenamefont
  {Jackson}, \citenamefont {M\"{o}ller},\ and\ \citenamefont
  {Roy}}]{jackson14}%
  \BibitemOpen
  \bibfield  {author} {\bibinfo {author} {\bibfnamefont {T.~S.}\ \bibnamefont
  {Jackson}}, \bibinfo {author} {\bibfnamefont {G.}~\bibnamefont {M\"{o}ller}},
  \ and\ \bibinfo {author} {\bibfnamefont {R.}~\bibnamefont {Roy}},\
  }\href@noop {} {\bibfield  {journal} {\bibinfo  {journal} {Nat. Comm.}\
  }\textbf {\bibinfo {volume} {6}},\ \bibinfo {pages} {8629} (\bibinfo {year}
  {2014})}\BibitemShut {NoStop}%
\bibitem [{\citenamefont {Lee}\ \emph {et~al.}(2017)\citenamefont {Lee},
  \citenamefont {Claassen},\ and\ \citenamefont {Thomale}}]{lee17}%
  \BibitemOpen
  \bibfield  {author} {\bibinfo {author} {\bibfnamefont {C.~H.}\ \bibnamefont
  {Lee}}, \bibinfo {author} {\bibfnamefont {M.}~\bibnamefont {Claassen}}, \
  and\ \bibinfo {author} {\bibfnamefont {R.}~\bibnamefont {Thomale}},\
  }\href@noop {} {\bibfield  {journal} {\bibinfo  {journal} {Phys. Rev. B}\
  }\textbf {\bibinfo {volume} {96}},\ \bibinfo {pages} {165150} (\bibinfo
  {year} {2017})}\BibitemShut {NoStop}%
\bibitem [{\citenamefont {Mielke}(1991)}]{mielke91}%
  \BibitemOpen
  \bibfield  {author} {\bibinfo {author} {\bibfnamefont {A.}~\bibnamefont
  {Mielke}},\ }\href@noop {} {\bibfield  {journal} {\bibinfo  {journal} {J.
  Phys. A: Math. Gen.}\ }\textbf {\bibinfo {volume} {24}},\ \bibinfo {pages}
  {L73} (\bibinfo {year} {1991})}\BibitemShut {NoStop}%
\bibitem [{\citenamefont {Sondhi}\ \emph {et~al.}(1993)\citenamefont {Sondhi},
  \citenamefont {Karlhede}, \citenamefont {Kivelson},\ and\ \citenamefont
  {Rezayi}}]{sondhi93}%
  \BibitemOpen
  \bibfield  {author} {\bibinfo {author} {\bibfnamefont {S.~L.}\ \bibnamefont
  {Sondhi}}, \bibinfo {author} {\bibfnamefont {A.}~\bibnamefont {Karlhede}},
  \bibinfo {author} {\bibfnamefont {S.~A.}\ \bibnamefont {Kivelson}}, \ and\
  \bibinfo {author} {\bibfnamefont {E.~H.}\ \bibnamefont {Rezayi}},\
  }\href@noop {} {\bibfield  {journal} {\bibinfo  {journal} {Phys. Rev. B}\
  }\textbf {\bibinfo {volume} {47}},\ \bibinfo {pages} {16419} (\bibinfo {year}
  {1993})}\BibitemShut {NoStop}%
\bibitem [{\citenamefont {Neupert}\ \emph {et~al.}(2011)\citenamefont
  {Neupert}, \citenamefont {Santos}, \citenamefont {Ryu}, \citenamefont
  {Chamon},\ and\ \citenamefont {Mudry}}]{neupert11}%
  \BibitemOpen
  \bibfield  {author} {\bibinfo {author} {\bibfnamefont {T.}~\bibnamefont
  {Neupert}}, \bibinfo {author} {\bibfnamefont {L.}~\bibnamefont {Santos}},
  \bibinfo {author} {\bibfnamefont {S.}~\bibnamefont {Ryu}}, \bibinfo {author}
  {\bibfnamefont {C.}~\bibnamefont {Chamon}}, \ and\ \bibinfo {author}
  {\bibfnamefont {C.}~\bibnamefont {Mudry}},\ }\href@noop {} {\bibfield
  {journal} {\bibinfo  {journal} {Phys. Rev. Lett.}\ }\textbf {\bibinfo
  {volume} {108}},\ \bibinfo {pages} {046806} (\bibinfo {year}
  {2011})}\BibitemShut {NoStop}%
\bibitem [{\citenamefont {Bergholtz}\ and\ \citenamefont
  {Liu}(2013)}]{bergholtz13}%
  \BibitemOpen
  \bibfield  {author} {\bibinfo {author} {\bibfnamefont {E.~J.}\ \bibnamefont
  {Bergholtz}}\ and\ \bibinfo {author} {\bibfnamefont {Z.}~\bibnamefont
  {Liu}},\ }\href@noop {} {\bibfield  {journal} {\bibinfo  {journal} {Int. J.
  Mod. Phys. B}\ }\textbf {\bibinfo {volume} {27}},\ \bibinfo {pages} {1330017}
  (\bibinfo {year} {2013})}\BibitemShut {NoStop}%
\bibitem [{\citenamefont {Regnault}\ and\ \citenamefont
  {Bernevig}(2011)}]{regnault11}%
  \BibitemOpen
  \bibfield  {author} {\bibinfo {author} {\bibfnamefont {N.}~\bibnamefont
  {Regnault}}\ and\ \bibinfo {author} {\bibfnamefont {B.~A.}\ \bibnamefont
  {Bernevig}},\ }\href@noop {} {\bibfield  {journal} {\bibinfo  {journal}
  {Phys. Rev. X}\ }\textbf {\bibinfo {volume} {1}},\ \bibinfo {pages} {021014}
  (\bibinfo {year} {2011})}\BibitemShut {NoStop}%
\bibitem [{\citenamefont {Kourtis}\ \emph {et~al.}(2014)\citenamefont
  {Kourtis}, \citenamefont {Neupert}, \citenamefont {Chamon},\ and\
  \citenamefont {Mudry}}]{kourtis14}%
  \BibitemOpen
  \bibfield  {author} {\bibinfo {author} {\bibfnamefont {S.}~\bibnamefont
  {Kourtis}}, \bibinfo {author} {\bibfnamefont {T.}~\bibnamefont {Neupert}},
  \bibinfo {author} {\bibfnamefont {C.}~\bibnamefont {Chamon}}, \ and\ \bibinfo
  {author} {\bibfnamefont {C.}~\bibnamefont {Mudry}},\ }\href@noop {}
  {\bibfield  {journal} {\bibinfo  {journal} {Phys. Rev. Lett.}\ }\textbf
  {\bibinfo {volume} {112}},\ \bibinfo {pages} {126806} (\bibinfo {year}
  {2014})}\BibitemShut {NoStop}%
\bibitem [{\citenamefont {Throckmorton}\ and\ \citenamefont
  {Vafek}(2011)}]{throckmorton11}%
  \BibitemOpen
  \bibfield  {author} {\bibinfo {author} {\bibfnamefont {R.~E.}\ \bibnamefont
  {Throckmorton}}\ and\ \bibinfo {author} {\bibfnamefont {O.}~\bibnamefont
  {Vafek}},\ }\href@noop {} {\bibfield  {journal} {\bibinfo  {journal} {Phys.
  Rev. B}\ }\textbf {\bibinfo {volume} {86}},\ \bibinfo {pages} {115447}
  (\bibinfo {year} {2011})}\BibitemShut {NoStop}%
\bibitem [{\citenamefont {Pizarro}\ \emph {et~al.}(2019)\citenamefont
  {Pizarro}, \citenamefont {R\"osner}, \citenamefont {Thomale}, \citenamefont
  {Valent\'{\i}},\ and\ \citenamefont {Wehling}}]{pizarro19}%
  \BibitemOpen
  \bibfield  {author} {\bibinfo {author} {\bibfnamefont {J.~M.}\ \bibnamefont
  {Pizarro}}, \bibinfo {author} {\bibfnamefont {M.}~\bibnamefont {R\"osner}},
  \bibinfo {author} {\bibfnamefont {R.}~\bibnamefont {Thomale}}, \bibinfo
  {author} {\bibfnamefont {R.}~\bibnamefont {Valent\'{\i}}}, \ and\ \bibinfo
  {author} {\bibfnamefont {T.~O.}\ \bibnamefont {Wehling}},\ }\href@noop {}
  {\bibfield  {journal} {\bibinfo  {journal} {Phys. Rev. B}\ }\textbf {\bibinfo
  {volume} {100}},\ \bibinfo {pages} {161102} (\bibinfo {year}
  {2019})}\BibitemShut {NoStop}%
\bibitem [{\citenamefont {Vanhala}\ and\ \citenamefont
  {Pollet}(2020)}]{vanhala2020}%
  \BibitemOpen
  \bibfield  {author} {\bibinfo {author} {\bibfnamefont {T.~I.}\ \bibnamefont
  {Vanhala}}\ and\ \bibinfo {author} {\bibfnamefont {L.}~\bibnamefont
  {Pollet}},\ }\href@noop {} {\bibfield  {journal} {\bibinfo  {journal} {Phys.
  Rev. B}\ }\textbf {\bibinfo {volume} {102}},\ \bibinfo {pages} {035154}
  (\bibinfo {year} {2020})}\BibitemShut {NoStop}%
\end{thebibliography}%
